\documentclass[useAMS,usenatbib]{mn2e}
\usepackage{graphicx}    
\usepackage{subfigure}
\usepackage[latin1]{inputenc}
\usepackage{amsmath}
\usepackage{amsfonts}
\usepackage{amssymb}
\usepackage{color}

\def \fps@figure{htbp}
\makeatother

\DeclareGraphicsExtensions{.ps,.pdf,.png}
\makeatletter

\def \re     {$R_{\rm e}$}
\def \sersic  {S\'{e}rsic}
\def \rff     {\textit{RFF}}

\begin{document}

\title[Evolution of BCGs since $z\sim 2$]{Exploring the progenitors of brightest cluster galaxies at $z\sim 2$}

\author[Zhao et al.]{Dongyao Zhao$^{1}$\thanks{E-mail: : \texttt{ppxdz1@nottingham.ac.uk}}, Christopher~J.~Conselice$^{1}$, Alfonso Arag\'{o}n-Salamanca$^{1}$, \newauthor Omar Almaini$^{1}$, William G. Hartley$^{1,2}$, Caterina Lani$^{1,3}$, Alice Mortlock$^{1,4}$, \newauthor Lyndsay Old$^{1}$\\
\\$^{1}$School of Physics and Astronomy, The University of Nottingham, University Park, Nottingham, NG7 2RD, UK
\\$^{2}$ETH Z{\"u}rich, Institut f{\"u}r Astronomie, Wolfgang-Pauli-Str. 27, CH-8093 Z{\"u}rich, Schweiz
\\$^{3}$School of Physics and Astronomy and the Wise Observatory, The Raymond and Beverly Sackler Faculty of Exact Sciences,\\ Tel-Aviv University, Tel-Aviv 69978, Israel
\\$^{4}$SUPA\thanks{Scottish Universities Physics Alliance} Institute for Astronomy, The University of Edinburgh, Royal Observatory, Edinburgh, EH9 3HJ, UK }


\date{Accepted 20th September 2016. Received 18th September 2016; in original form 29th April 2016}
\pagerange{\pageref{firstpage}--\pageref{lastpage}} \pubyear{2014}
\maketitle

\label{firstpage}

\begin{abstract}
We present a new method for tracing the evolution of BCGs from $z\sim 2$ to $z\sim 0$. We conclude on the basis of semi-analytical models that the best method to select BCG progenitors at $z\sim 2$ is a hybrid environmental density and stellar mass ranking approach. Ultimately we are able to retrieve 45\% of BCG progenitors. We apply this method on the CANDELS UDS data to construct a progenitor sample at high redshift. We furthermore populate the comparisons in local universe by using SDSS data with statistically likely contamination to ensure a fair comparison between high and low redshifts. Using these samples we demonstrate that the BCG sizes have grown by a factor of $\sim 3.2$ since $z\sim 2$, and BCG progenitors are mainly late-type galaxies, exhibiting less concentrated profiles than their early-type local counterparts. We find that BCG progenitors have more disturbed morphologies. In contrast, local BCGs have much smoother profiles. Moreover, we find that the stellar masses of BCGs have grown by a factor of $\sim 2.5$ since $z\sim 2$, and the SFR of BCG progenitors has a median value of 13.5 $M_\odot$yr$^{-1}$, much higher than their quiescent local descendants. We demonstrate that over $z=1-2$ star formation and merging contribute equally to BCG mass growth. However, merging plays a dominant role in BCG assembly at $z \lesssim 1$. We also find that BCG progenitors at high-$z$ are not significantly different from other galaxies of similar mass at the same epoch. This suggests that the processes which differentiate BCGs from normal massive elliptical galaxies must occur at $z \lesssim 2$.
\end{abstract}

\begin{keywords}
galaxies: clusters: general --- galaxies: evolution --- galaxies: formation
\end{keywords}

\section{Introduction} 

Brightest cluster galaxies (BCGs) are the most luminous and massive galaxies in local universe. They reside at the bottom of the gravitational potential well of galaxy clusters, and are surrounded by a population of satellite galaxies.  The special regions they reside in, and the unique properties they exhibit (e.g., distinct structures and morphologies, very high stellar masses) set them apart from the general galaxy population. Their origin and evolution also tightly link with the evolution of their host clusters and provide direct information on the history of large-scale structures in Universe (e.g., \citealt{Conroy07}). Even though much attention has been dedicated to the study of BCG formation and evolution, understanding when these most massive galaxies formed and how they evolve with time are still controversial issues.

Early N-body simulations studying BCG formation through merging in a cold matter (CDM) cosmology, find that BCG growth through early merging of few massive galaxies dominates over late-time accretion of many smaller systems (e.g., \citealt{Dubinski98}). The modern context of BCG assembly through hierarchical growth within networks of dark matter halos is now well established. For example,  by using nine high-resolution dark matter-only simulations of galaxy clusters in a $\Lambda$CDM universe, \citet{Laporte13} claim that BCGs can grow mainly through dissipationless dry mergers of quiescent galaxies from $z = 2$ to the present day, producing BCG light profiles and stellar mass growth in good agreement with observations. However, pure N-body models ignore mechanisms such as gas cooling and star formation in BCG evolution which are also likely important processes.  

Taking into account hydrodynamical processes such as infalling gas and AGN feedback, recent semi-analytic models (SAMs) suggest that the stellar component of  today's BCGs was initially formed through the collapse of cooling gas or gas-rich mergers at high redshift, and consequently BCGs continued to grow, but assemble substantially very late ($50$\% of the final mass is assembled at $z \lesssim 0.5$) through dissipationless processes such as dry mergers of satellite galaxies (\citealt{DeLB07}; \citealt{Naab09}; \citealt{Laporte12}). This two-phase evolution for BCG growth  successfully reproduces many observations, however, it has been questioned by a number of studies which find a much slower stellar mass growth in BCGs at $z\lesssim 1$ in observations  (e.g., \citealt{Whiley08}; \citealt{Collins09}; \citealt{Lin13}; \citealt{Zhang15}). More observational studies of BCGs at higher redshifts will help to constrain these models and give us a better idea of their evolution.

To understand how BCGs evolved and assembled their stellar masses, and which mechanisms drive these changes, it is important to properly connect today's BCGs to their progenitors at earlier times observationally. This requires the non-trivial task of linking BCG descendants with their progenitors through cosmic time, which in turn requires assumptions for how BCGs evolve. 
 
At lower redshift ($z\lesssim 1-1.5$), BCG progenitor-descendant pairs are selected by an empirical approach through constructing a sample  based on finding distant clusters, and using the correlation between BCG stellar mass and cluster mass. Employing this method, many studies have  characterized the assembly of BCGs at $z \lesssim 1$. \citet{Lidman12} demonstrated that BCGs have grown by a factor of $1.8$ between $z = 0.2-0.9$. While \citet{Lin13} found a similar growth such that the stellar mass of BCGs increases by a factor of $\sim 2.3$ since $z \sim 1.4$. \citet{Shankar15} claimed an increase of a factor $\sim 2-3$ in BCG mean stellar mass, and $\sim 2.5-4$ factor increase in BCG mean effective radius, since $z \sim 1$. \citet{Zhang15} showed a BCGs mass growth by a factor of $\sim 2$ since $z \sim 1.2$ using a similar approach.

However, the techniques  for linking local BCGs and their progenitors at  $z \lesssim 1$ are difficult to apply at higher redshifts ($z \gtrsim 1.5$). On the one hand, it is difficult to identify clusters/proto-clusters at early times. On the other hand, it is also difficult to define BCG progenitors in high-$z$ clusters since the main progenitor may not be the most luminous/massive galaxy  as the low-z BCGs.  Nonetheless, a number of studies have been carried out to explore the build-up of massive galaxies up to $z\sim 3$.

Among the solutions for  linking galaxies at different redshifts, matching galaxy progenitors and descendants  at a constant number density has been demonstrated to be a considerably improved approach for tracking the evolution of galaxies (e.g., \citealt{Leja13}; \citealt{Mundy15}). By applying this method, \citet{vanDokkum10} claim a mass growth of a factor of $\sim 2$, and a size growth of a factor of $\sim 4$ for massive galaxies since $z = 2$. \citet{Ownsworth14}, using a variety of number density selections with $n\leqslant 1\times10^{-4}$Mpc$^{-3}$ at $0.3 < z < 3$, find that about 75\% of the total stellar mass in massive galaxies at $z = 0.3$ is created at $z < 3$, and the sizes of massive galaxy progenitors is a factor of 1.8 smaller than local early-type galaxies of similar mass. \citet{Marchesini14} investigate ultra-massive galaxy evolution by using progenitors from $z = 3$ which are selected with both a fixed cumulative number density and an evolving number density. They find that the stellar content of ultra-massive galaxies have grown by a factor of $2-3.6$ since $z =3$.  However, these systems are not necessarily BCGs, and a clear correspondence between massive galaxies and BCGs at high redshifts ($z \gtrsim 1.5$) is still lacking. In order to obtain better perspective of BCG assembly, it is critical to identify the progenitors of BCGs at $z \gtrsim 1.5$, and to explore their mass and structural evolution.    


Mergers are potentially a significant process in BCG formation, as they are predicted to be a major mechanism in the hierarchical picture of galaxy formation. Apart from the dominant role of minor mergers in BCG mass assembly at low redshift (e.g., \citealt{Burke15}), observations suggest that  at high redshifts  BCG evolution is also largely driven by mergers through both major and minor events (e.g., \citealt{Lidman13}; \citealt{Burke13}). Since mergers closely relate to the environmental density around galaxies, in this paper, we propose a method to identify BCG progenitors at $z \sim 2$, which depends on galaxy local densities as well as galaxy stellar masses. We first examine  the effectiveness of our method using simulation data, and then apply this method on the observational data of the CANDELS UDS survey.  Our method to probe BCG progenitors at $z \gtrsim 1.5$ is easier, since it avoids the difficulty of identifying clusters at high redshifts. Comparing high-$z$ BCG progenitors with their local SDSS descendants, we  study the evolution of BCG structure, morphology, stellar mass and star formation since $z\sim 2$, and discuss the implied formation processes for BCGs.

The rest of this paper is organized as follows. In Section ~\ref{sec:data_and_technique}, we present the observational data employed in this work. We also introduce necessary quantities which will be used in selecting our BCG progenitors and for comparing BCG properties in this section. The description and simulation tests of our selection of BCG progenitors are presented in Section~\ref{sec:progselect}. Although the BCG progenitors selected by our method are contaminated by non-BCG progenitors, in Section~\ref{sec:contam_effect}, we demonstrate that our selected progenitors sample, and their local descendants, can be used to trace BCG evolution since $z\sim 2$. We then describe our results of BCG assembly in Section~\ref{sec:results}.  In Section~\ref{sec:discussion} we first discuss the possible mechanisms for BCG evolution implied by our results, and then we compare our results with other studies of BCG evolution at $z \lesssim 1$ as well as massive galaxy growth since $z\sim 2$. Finally, we summarise our results in Section~\ref{sec:summary}. Throughout this paper we have adopted the $\Lambda$CDM cosmology with $\Omega_m=0.3$, $\Omega_\Lambda=0.7$, and $H_0=70$ km s$^{-1}$ Mpc$^{-1}$.

\section{Observational Data and Quantities}
\label{sec:data_and_technique}


In this section, we first describe the galaxy catalogues for our local samples and high-$z$ galaxies. We also provide information on their properties, such as stellar masses, star formation rates (SFRs) and specific star formation rates (sSFRs). The environmental density measured at high redshifts in observations is introduced in a following separate subsection. We then explain how we use the constant number density of  galaxy environment  to trace BCG evolution. The structural properties of galaxies from profile fitting are described in the final subsection. 

\subsection{Local Sample}
The local BCG sample in this paper, which we compare with the high-$z$ progenitors to study BCG evolution, comes from the catalogue published by \citet{Linden07} (hereafter is L07). The groups and clusters that host these BCGs are found in the Sloan Digital Sky Survey (SDSS) based C4 cluster catalogue (\citealt{Miller05}), a widely used and well-defined sample whose reliability has been thoroughly tested by simulations. Based on the C4 sample, L07 developed an improved algorithm to identify BCGs in clusters, and published a catalogue containing 625 BCGs residing in galaxy groups and clusters at $0.02 \leq z \leq 0.10$ (see L07 for a detailed discussion on the BCG identification method).


The stellar masses we use for the L07 BCGs are  the ``The MPA--JHU DR7 release of spectrum measurements'' (see http://www.mpa-garching.mpg.de/SDSS/DR7/)\footnote{In this paper we use their updated stellar masses from \texttt{http://home.strw.leidenuniv.nl/$\sim$jarle/SDSS/}}. The stellar mass for SDSS galaxies was initially derived by fitting the observed values of the $D_n(4000)$ and $H \delta_A$ indices with a library of models from \citet{BC03} (\citealt{Kauffmann03}). A Kroupa IMF is assumed. MPA--JHU group then used broad-band $u, g, r, i, z$ photometry of SDSS DR7 for the spectral energy distribution (SED) fits instead of the spectral features. Although the method is not identical to that of \citet{Kauffmann03}, the results agree very well. A detailed discussion and comparison of the methods can be found in http://mpa-garching.mpg.de/SDSS/DR7/mass\_comp.html. The MPA--JHU mass is the median of this distribution. Compared with the stellar mass obtained from the best $\chi^2$ model, median stellar mass is $\sim 0.1$ dex smaller (\citealt{Brinchmann04}; \citealt{Fernandes05}). When we compare the stellar mass between local sample and their high-$z$ progenitors, we convert the MPA--JHU stellar masses to those with a Chabrier IMF.

As we discuss in detail later in this paper, the high-$z$ progenitor sample selected by our method will contain both true BCG progenitors and non-BCG progenitors (see Section~\ref{sec:prog_fraction}).  This is due to there being no perfect way to only select true BCG progenitors at high redshift.   Thus, at $z\sim 0$, we  construct a counterpart sample which is a mixture of local BCGs and local non-BCGs as the descendants of our high-$z$ progenitors (see  Section~\ref{sec:nocontam_local}). Therefore, in addition to the L07 BCG catalogue, we also employ SDSS DR7 data as our parent galaxy catalogue to select local non-BCGs to match the high-$z$ inevitable contamination. Since the non-BCGs are selected based on their stellar mass (see Section~\ref{sec:nocontam_local}), the parent galaxy catalogue  is the MPA--JHU DR7 stellar mass catalogues. We select our ``contamination'' galaxies within the redshift range of $0.02 \leq z \leq 0.10$, the same as our BCG sample.  


The SFR and sSFR for both the pure BCGs and the contaminant non-BCGs are taken from the MPA--JHU SFR catalogue (http://mpa-garching.mpg.de/SDSS/DR7/sfrs.html). The total SFRs (dust-corrected) for star-forming galaxies are derived by \citet{Brinchmann04} based on a H$\alpha$ emission line modelling technique.  \citet{Salim07} demonstrated that these ``H$\alpha$'' SFRs are very consistent with the dust-corrected SFRs constrained by the UV luminosity of local star-forming galaxies. For local galaxies without H$\alpha$ detections which belong almost exclusively in the red sequence, the dust-corrected SFRs are obtained from SED fitting of SDSS photometry (details could be found in \citealt{Salim07}). The SFRs for SDSS galaxies are measured by assuming a Kroupa IMF. They are divided by $1.06$ when compared with the SFRs of high-$z$ galaxies which are derived by assuming a Chabrier IMF. $1.06$ is the conversion factor to convert SFRs which are calculated for Kroupa IMF to Chabrier IMF. sSFRs of SDSS galaxies are calculated by using the SFRs described here and the MPA--JHU masses. When compared with high-$z$ sSFR, they are also converted to the values for Chabrier IMF.

\subsection{High-$z$ Sample}

The Cosmic Assembly Near-infrared Deep Extragalactic Legacy Survey (CANDELS; PIs: Faber and Ferguson; \citealt{Grogin11}; \citealt{Koekemoer11}) provides excellent data to study galaxy properties at high redshift. CANDELS is a 902-orbit Multi-Cycle Treasury program on the \textit{Hubble Space Telescope (HST)} with imaging by the Wide Field Camera 3 (WFC3) and the Advanced Camera for Surveys (ACS) on five different fields: GOODS-N, GOODS-S, COSMOS and UDS. The galaxy catalogue on which we apply our selection of BCG progenitors is from the CANDELS UDS (\citealt{Mortlock15}). 
 
CANDELS UDS covers a part of the field of UKIRT Infrared Deep Sky Survey (UKIDSS, \citealt{Lawrence07}) Ultra Deep Survey (UDS). Its image has a pixel scale of 0.06 arcsec/pixel and  a $5\sigma$ depth of $H=26.3$ in a 1 arcsec aperture. The photometry of the CANDELS UDS includes \textit{U}-band data from the CFHT (Foucaud et al. in prep), \textit{B, V, R, i', z'}-band data from the Subaru/XMM-Newton Deep Survey (SXDS; \citealt{Furusawa08}), \textit{J, H} and \textit{K}-band data from UKIDSS UDS, \textit{F}606\textit{W} and \textit{F}814\textit{W} data from the ACS, $H_{160}$ and $J_{125}$-band WFC3 data, \textit{Y} and \textit{Ks} bands taken as part of the Hawk-I UDS and GOODS Survey (HUGS; VLT large programme ID 186.A-0898, PI: Fontana; \citealt{Fontana14}).

The photometric redshifts for the high-$z$ galaxies of CANDELS UDS are calculated by \citet{Mortlock15} with the method descried in \citet{Hartley13}. In brief, the SED templates are fit to the photometry described above, and the best-fitting redshift is used. We constrain our high-$z$ galaxy sample within the redshift range of $1\leq z\leq 3$, to ensure a statistically large number of high-$z$ progenitors selected by our method. 


The stellar masses of our high-$z$ galaxies are calculated by \citet{Mortlock15} for the CANDELS UDS. The method used to compute the stellar masses is described in detail in \citet{Mortlock13, Mortlock15}, \citet{Hartley13} and \citet{Lani13}. Briefly, the stellar masses are measured through a multi-colour stellar population fitting technique.  With a Chabrier IMF, a large grid of synthetic SEDs from the stellar population models of \citet{BC03} are used to fit the multi-band photometry of the CANDELS UDS. They obtained two kinds of stellar mass. One is the best-fit stellar mass whose template has the smallest $\chi^2$ value. Another one is the mode stellar mass. By binning the stellar masses of the 10\% of templates with the lowest $\chi^2$ in bins of 0.05 dex, they determine the mode stellar mass which corresponds to the stellar mass bin with the largest number of templates. In this work we use the mode stellar masses in the catalogue as these masses are less likely to be affected by the bad fitting through templates (\citealt{Mortlock13}). 

We find that for all the CANDELS UDS galaxies, the mode stellar mass is statistically consistent with the best-fit stellar mass. Mode mass is only $\sim 0.01$ dex smaller than the best-fit one. For our selected high-$z$ progenitors (selection is described in Section~\ref{sec:progselect}) which are more massive, we find that the difference between mode and best-fit mass becomes larger, such that the best-fit stellar mass is $\sim 0.1$ dex larger than the mode mass. Note that in local universe the best-fit mass for SDSS galaxies is $\sim 0.1$ dex larger than the MPA--JHU mass. Although the methods used to determine stellar masses at low and high redshift are not exactly the same,  the principles applied--SED fitting of rest-frame optical data--are very similar. Moreover, given the evident BCG mass growth we measure (a factor of $\sim2.5$; see Section~\ref{sec:smass_growth}), it is not unreasonable to assume that such a large effect cannot be solely explained by systematic differences in the stellar mass determination.

Since CANDELS UDS is a subset of the UDS field, it benefits from the same wealth of the UDS data set, such as the SFR. The SFRs we use for our high-$z$ galaxies are calculated by \citet{Ownsworth14} for the full UDS field. They are obtained from the rest-frame near UV luminosities which trace the presence of young and short-lived stellar populations produced by recent star formation. First, \citet{Ownsworth14} determine dust-uncorrected SFRs with a Chabrier IMF. Since the UV light is very susceptible to dust extinction, they then apply a careful dust correction to obtain the final dust-corrected SFRs. For the full description of the dust correction and SFR calculation see \citet{Ownsworth14}. The sSFRs are calculated by taking these SFRs and the stellar masses described above.

In Section~\ref{sec:SFR_sSFR}, we find that BCG progenitors at $z \sim 2$ have much higher SFRs (by almost two orders of magnitude) than their local quiescent descendants. Although the techniques for measuring SFRs of high-$z$ and local galaxies are not exactly the same, since the H$\alpha$ and UV SFRs are very similar (e.g., \citealt{Salim07}; \citealt{Twite12}) and all SFRs are carefully dust-corrected, we think our statistical results are still reliable and reasonably robust. Explicitly,  we do not think the uncertainty introduced by the different SFR measurements used at high- and low-$z$ are responsible for the clear SFR evolution that we detect.


\subsection{Density Measurement in Observations}
\label{sec:density_obs}

One important property we use in this work to select the BCG progenitors is the local environmental density around the high-$z$ galaxies. \citet{Lani13} compute the environmental density for UDS galaxies which can be also used for the CANDELS UDS sample. The detailed discussion of the density measurement can be found in their paper. In brief, the densities we use in this work are measured by galaxy counts in a fixed physical aperture. 

\citet{Lani13} construct a cylinder with a projected radius of 400 kpc and depth of 1 Gyr around each galaxy within which they count the number of neighbouring galaxies. The radius of 400 kpc represents the typical ``radius'' of galaxy clusters at high redshifts. The depth of 1 Gyr is several times greater than the 1$\sigma$ measured uncertainty on the photometric redshifts. This depth avoid diluting the number of galaxies in the cylinder by minimising the exclusion of sources due to the large photometric redshift errors. Moreover, with accounting for holes and edges in the field, the number of real galaxies in an aperture ($N_{\rm g}^{\rm aper}$) is normalised. The equation to calculate the density for every galaxy in the UDS catalogue is

\begin{equation}
\label{eq:density_UDS}
\rho_{\rm aper} = \frac{N_{\rm g}^{\rm aper}}{N_{\rm mask}^{\rm aper}} \times \frac{N_{\rm mask}^{\rm tot}}{N_{\rm z}},
\end{equation}

\noindent where $N_{\rm mask}^{\rm aper}$ is the number of good pixels which are not masked within the chosen aperture, $N_{\rm mask}^{\rm tot}$ is the total number of non-masked pixels in the UDS, and $N_{\rm z}$ is the total number of galaxies over the entire field which lie within the 1 Gyr redshift interval we consider. 

The galaxies employed in \citet{Lani13} to calculate the environmental density are taken from the UDS K-band selected catalogue. They applied a magnitude completeness cut of $K_{\rm AB} = 24.4$, which produces a completeness of $\sim$99\%. This magnitude cut corresponds to a stellar mass limit of $M_{*}^{\rm cut}=10^{9.76} M_\odot$ at $z \sim 2$, assuming a Chabrier IMF. All galaxies at $z \sim 2$ with stellar masses above $M_{*}^{\rm cut}=10^{9.76} M_\odot$ are included in this K-band selection. For more details we refer the reader to \citet{Hartley13} and \citet{Mortlock15}.

\subsection{Constant Number Density Selection} 
\label{sec:constant_ND}

In this section we discuss how to connect our low redshift sample of BCGs to the galaxies at high redshifts.
 
The main method we use to identify BCG progenitors and study BCG evolution is to match the abundance of BCG environments at low and high redshift. In other words, we  assume a constant number density of ``BCG environments'' (i.e., a constant number density of clusters or highest-density regions). Since BCGs reside in some of the densest environments and are hosted by the most massive halos in the local universe, it is reasonable to assume that at high-$z$ each BCG progenitor also likely reside in one of the most overdense regions. Each high-$z$ overdensity hosting the BCG progenitor may accrete galaxies from other less dense regions, and finally evolve into one galaxy cluster hosting a BCG in the local universe. We assume that the comoving number density of local galaxy clusters and that of the high-$z$ most overdense regions that host the BCG progenitors are approximately the same, with consideration that mergers among the most massive clusters/halos hosting BCGs or BCG progenitors are expected to be much rarer than among normal galaxies (or much less massive halos). 

\citet{FakhouriMaBoylan10}  find in simulations that the merger rate per dark matter halo is nearly independent of halo mass, so that the major/minor mergers are as common among massive halos as they are among less massive ones. Thus, taking into account the merger of BCG environments, we consider  the effect of using an evolving number density of BCG environments in our selection of BCG progenitors. In our simulations we find that for one BCG at $z=0$ there are, on average, $1.4$ overdensities at $z \sim 2$ whose most massive galaxies will end up in it.  Therefore, applying an evolving environment number density of $1.4\times10^{-4.06}h^3\,$Mpc$^{3}$ at the $z = 2.07$ snapshot in the simulation (i.e., 1.4 times larger than the non-evolving one), we find that the fraction of true BCG progenitors in the selected sample is comparable (actually, marginally smaller) than the one found using constant number density. Therefore, using an evolving number density does not improve the success rate of the BCG progenitor selection; on the contrary, the sample is contaminated by a slightly higher fraction of non-BCG progenitors. Furthermore, translating an evolving number density of structures from the simulations into the observational domain at $z\sim2$ is likely to introduce further uncertainties. Since the additional complications inherent in considering evolving number densities do not seem to improve the results, we opt for the simpler constant number density in our environmental matching method.


The  number density of environment used in our study corresponds to that of the clusters in L07. We consider the clusters whose velocity dispersions are $\sigma_{200} \geq 309$ km/s, with DM halo masses $M_{200} \geq 10^{13.55} h^{-1} M_\odot$, corresponding to a cumulative comoving number density of  $10^{-4.06} h^3$Mpc$^{-3}$. Selecting the most overdense regions down to this number density limit by using the CANDELS UDS data, we need to select $38$ the densest environments at $1 \leq z \leq 3$. We then identify the BCG progenitor  as the most massive galaxy in each high-$z$ overdensity. Finally, there are $38$ BCG progenitors selected from the CANDELS UDS. Correspongding to the comoving number density of  $10^{-4.06} h^3$Mpc$^{-3}$, $469$ galaxies are selected at $0.02 \leq z \leq 0.10$ from SDSS DR7 as the local comparisons. Detailed descriptions on how we choose our $38$ high-$z$ and $469$ local sample  are presented in Section~\ref{sec:basic_idea} and Section~\ref{sec:nocontam_local}.  

Note that by using this method, not all of the selected massive galaxies are true BCG progenitors. In Section~\ref{sec:progselect} we will look at the fraction of true BCG progenitors in the selected progenitor sample obtained by our method, using simulation data. We also examine the fraction of true BCG progenitors in the selected progenitor sample obtained by using a fixed galaxy number density (the method normally used when studying massive galaxy evolution, see \citealt{vanDokkum10} and \citealt{Ownsworth14}). Comparing these methods, we conclude  that our environmental matching is a better way to identify true BCG progenitors at high redshifts. Our main results are therefore obtained using this method.

\subsection{Shifting Local Galaxies to High Redshift}  

\begin{figure}
\center{\includegraphics[scale=0.7]{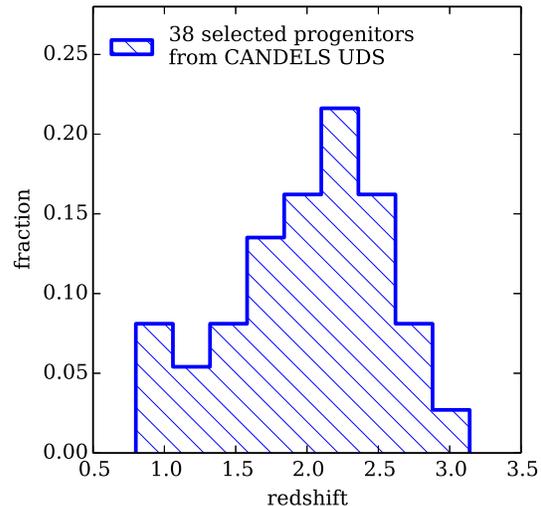}}
\caption{Redshift distribution of the 38 progenitors selected by our method as the most massive galaxies in the densest environments from CANDELS UDS. The median redshift of this distribution is $z=2.06$.}  
\label{fig:candelsz}
\end{figure}

\begin{figure*}
\center{\includegraphics[scale=0.53]{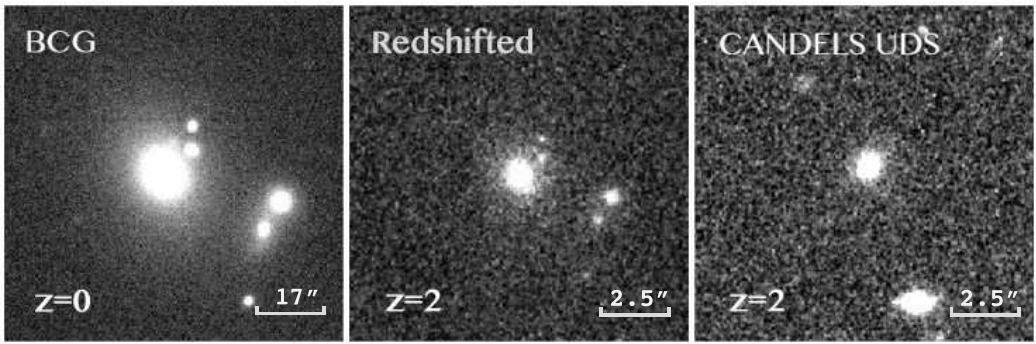}}
\caption{An example of a simulated galaxy created by using the FERENGI code (middle panel) after shifting one local BCG in the SDSS $g$-band (left panel) to $z=2$ as observed in the CANDELS UDS $H_{160}$-band data. The right panel is an original $H_{160}$-band image of a random galaxy at $z\sim 2$. This shows that the input we use in the FERENGI code is able to create a reasonable simulated image compared to an actual $z \sim 2$ image within the CANDELS 
$H_{160}$-band.}
\label{fig:shiftimage}
\end{figure*}

One aspect of BCG evolution we study in detail is the connection between local BCGs and their high-$z$ progenitors based on their structural evolution. The high spatial resolution and the high-quality images of the CANDELS UDS data allow for a good assessment of the structural properties (e.g., galaxy size and shape) of high-$z$ galaxies. However, a given galaxy will look different when observed with different instruments or at different redshifts. The extracted structural parameters are also wavelength dependent due to
bandpass shifting and cosmological dimming. Therefore, a direct comparison of structural parameters  from the original SDSS images and the CANDELS UDS images cannot be done without understanding these biases.

In order to explore the intrinsic structural evolution of BCGs, the images from the SDSS and CANDELS UDS need to be calibrated to allow comparisons between redshifts, ensuring similar resolutions and imaging depth. This can be achieved by using the code FERENGI (Full and Efficient Redshifting of Ensembles of Nearby Galaxy Images; \citealt{Barden08}). This code takes into account the cosmological corrections for size, surface brightness and bandpass shifting when simulating low redshift galaxies to high redshift. Simulated images are produced when the input galaxy images are simulated to appear as higher redshift images using the output redshift and instrumental properties.  For a full description about the code see \citet{Barden08}.

By applying our proposed BCG progenitor selection (detailed description in Section~\ref{sec:progselect}) on the CANDELS UDS data, the selected $38$ progenitors at $z=[1,3]$ have a redshift distribution as shown in Fig.~\ref{fig:candelsz}. To compare with this, the SDSS images therefore need to be simulated to $z=[1,3]$ following a similar redshift distribution shown in Fig.~\ref{fig:candelsz} after taking into account the $k$-correction in the FERENGI code. To be efficient when running FERENGI, we only simulate SDSS $g$-band images to CANDELS UDS $H_{160}$-band at $z=2$. This also allows us to account for the major $k$-correction because the $g$-band at $z \sim 0$ is in the same rest-frame wavelength as the $H_{160}$-band at $z=2$. 

Fig.~\ref{fig:candelsz} illustrates that $z \sim 2$ is the median redshift of our $38$ selected high-$z$ progenitors, and $\sim 90$\% of them are at $z < 2.5$, implying that the $k$-correction differences are not  a significant factor in the simulation.  Furthermore, we know that high-$z$ galaxies look very similar at wavelengths which are greater than the Balmer break (e.g., \citealt{Conselice11}) which is the case for our entire sample.   Testing on a small number of galaxies, we find that morphologies of the simulated SDSS galaxies placed at $z=2$ look very similar to the high-$z$ galaxies. We further demonstrate that galaxy structures (shape and size) measured from simulated images placed at $z=2$ do not have a large differential from the structures of their original galaxies.  We therefore only simulate SDSS $g$-band images to CANDELS UDS $H_{160}$-band at $z=2$ without a full $k$-correction.

One important input in the simulation code is the high redshift sky background image, whose size needs to be larger than the local input images. The size of our input SDSS galaxy images is $500 \times 500$ pixels which is too large to cut out a corresponding clean sky area within the CANDELS UDS image. Therefore we create simulated CANDELS UDS sky images which are large enough to be applied in the FERENGI code. First, we randomly choose 10 clean sky areas within the CANDELS imaging that contain no bright objects nearby. Within each of the sky areas, a patch of size  $200 \times 200$ pixels is cut out. Then for each patch we create the simulated sky image in $1000 \times 1000$ pixels by copying and pasting the patch. Ultimately, we create 10 simulated CANDELS UDS sky images for these simulations. Each of the SDSS galaxies are then redshifted within one of the simulated sky images which is randomly chosen from the ten.   
  
Since the stellar populations in galaxies at higher redshift are brighter and younger, simply shifting the local galaxies out to high redshift without considering the brightness increase due to stellar evolution will make them look fainter compared to the real average galaxies at such distances. In the FERENGI code, a brightness evolution is put in as an option to account for this evolution. It is introduced by a crude mechanism such that the magnitude evolves as $M_{\rm evo} = x \times z+M$. By studying the luminosity function from present to $z=2$, \citet{Ilbert05} found that the characteristic magnitude $M^*$ of the Schechter function in $B$ rest-frame band strongly evolves with redshift, such that $M^*$ at $z=2$ is $\sim 2$ magnitude smaller than that in local Universe. Since the SDSS $g$-band is similar with the $B$ band in rest-frame, we set $x=-1$, making a galaxy 2 mag brighter at redshift $z=2$ than it would be without luminosity evolution. 

The middle panel of Fig.~\ref{fig:shiftimage} shows one example of the output image from the FERENGI code after redshifting one local BCG in the SDSS $g$-band (left panel) to $z=2$ observed in CANDELS UDS $H_{160}$-band. The far right panel is the original $H_{160}$-band image of one random CANDELS UDS galaxy at $z=2$.  This demonstrates that the input we use in the FERENGI code is able to create a reasonable simulated image which appears similar to galaxies seen in the original $H_{160}$-band image at $z \sim 2$.

\subsection{Quantitative Characterisation of Galaxy Structure}

The surface brightness profiles of galaxies provide valuable information on their structure and their morphology. In addition to measuring  galaxy structural parameters by light profile fitting, we also introduce a parameter called the residual flux fraction (\rff ; \citealt{Hoyos11}) to quantify how good the model fit is and how far the galaxy profile deviates from the model profile.
 
\subsubsection{Structure Parameters}
The structural properties (effective radius \re\ and \sersic\ index $n$) of simulated local galaxies and high-$z$ progenitors are measured using 2D single \sersic \ (\citealt{Sersic63}) model fits. The \sersic\ model has the form 
\begin{equation}
\label{eq:sersic}
I(r)=I_{\rm e} \exp \{-b[(r/r_{\rm e})^{1/n}-1]\},
\end{equation}
where $I(r)$ is the intensity at distance $r$ from the centre, \re, the effective radius, is the radius that encloses half of the total luminosity, $I_e$ is the intensity at \re, $n$ is the \sersic \ index representing concentration, and $b \simeq 2n-0.33$ (\citealt{Caon93}). The fits are carried out with GALFIT (\citealt{Peng02}) through the  GALAPAGOS pipline (\citealt{Barden12}) in which the target galaxy and its near neighbours are fitted simultaneously, yielding more accurate results.
  
For local SDSS galaxies, these fits are also carried out on their $z=2$ simulated images created by the FERENGI code. For each target galaxy, the background level is fixed in GALFIT which is the mean sky value of the created CANDELS sky image used in the image simulation. The point spread function (PSF) employed in GALFIT is the output simulated PSF created by the FERENGI code.

For the high-$z$ galaxies we study in the CANDELS UDS, the structural parameters are measured from the HST WFC2 $H_{160}$ images with the PSF of this band, and with the sky value measured by GALAPAGOS. GALAPAGOS uses a flux growth curve method to improve the sky subtraction and produces a highly reliable measure of the background for single-band fits (\citealt{Haussler07}).  

\subsubsection{Residual Flux Fraction}
\label{sec:RFF_cal}
The light profiles of real galaxies are often complicated with features such as disturbances, merger remnants,  or other structures such as star-forming regions and spiral arms which cannot be fitted by a single \sersic\ model. Although we can do visual inspection on the residual images which will give us a good idea whether the galaxy profile can be explained by the single \sersic \ model, a more quantitative, repeatable, and objective diagnostic is desired to quantify how large the offset is after subtracting the single \sersic \ model from the original image. The \rff \ provides one such diagnostic, defined as 

\begin{equation}
\label{eq:rff}
RFF=\dfrac{\sum_{i,j\in A} |I_{i,j}-I_{i,j}^{\rm model}|-0.8\times \Sigma_{i,j\in A}\sigma_{i,j}^{\rm bkg}}{\Sigma_{i,j\in A} I_{i,j}},
\end{equation}

\noindent where $A$ is the particular aperture used to calculate the \rff , within which $I_{i,j}$ is the original flux of pixel ($i,j$), $I_{i,j}^{\rm model}$ is the model flux created by GALFIT, and $\sigma_{i,j}^{\rm bkg}$ is the \textit{rms} of the background light. The \rff \ measures the fraction of the signal contained in the residual that cannot be explained by background noise. See \citet{Hoyos11} for more details. 

The aperture $A$ we use to calculate the \rff \ is the ``Kron ellipse'', which is an ellipse with a semi-major axis of Kron radius\footnote{In this paper we use the following definition of ``Kron radius'': $R_{\rm kron}=2.5r_1$, where $r_1$ is the first moment of the light distribution \citep{Kron80,BA96}. For an elliptical light distribution, this is, strictly speaking, the semi-major axis.} ($R_{\rm kron}$) and the ellipticity and orientation determined by SExtractor for the galaxy. $\Sigma_{i,j \in A} I_{i,j}$ is computed as the total galaxy flux contained within the Kron ellipse, which is one of the SExtractor outputs, and therefore independent of the model fit. Additionally, to minimise  effects from nearby galaxies on the \rff , independently of whether they are fitted with the target galaxy simultaneously or not, we mask out the pixels belonging to all companions within the Kron ellipse using SExtractor segmentation maps. Therefore the \rff \ measures the residuals from the target galaxy fit alone. \citet{Zhao15a} includes a detailed discussion on the \rff \ calculation and how we apply it to our BCG samples.

We compute the \rff \ on the residual images of both the simulated local galaxies and the high-$z$ progenitors. The comparison of these will show at which epoch the galaxies are more disturbed, which we discuss later in this paper.  

\section{Selecting BCG Progenitors at $1<z<3$}
\label{sec:progselect}

In this section we introduce our basic procedure for the BCG progenitor selection.  This is a critical aspect and thus a major part of this paper.  Readers interested only in the comparison between the two redshifts can skip to Section~\ref{sec:results}.  In summary, we investigate how to match high-$z$ BCG progenitors with their $z = 0$ counterparts.  We ultimately selected a environmental matching method which depends on the  environments of galaxies at high redshifts to locate the most likely BCG progenitors. We test and fine-tune  our method using the output of the Millennium Simulation.

\subsection{Basic Assumption}
\label{sec:basic_idea}
In order to trace  the formation and evolution of BCGs, statistically large samples of BCGs are needed over a broad redshift range. In many other recent studies,  BCG samples at higher redshifts are selected through the detection of galaxy clusters in either the X-ray band (\citealt{Collins09}, \citealt{Burke13}, \citealt{Zhang15}) or the infrared band (\citealt{Lin13}), and  BCG evolution can be traced back to $z \sim 1$. Unfortunately, the observational constraint on  BCG evolutionary scenarios is still poor at $z \gtrsim 1-1.5$, and is limited by the difficulty of identifying large samples of galaxy clusters beyond $z \sim 1$. However, due to the fact that environment can be measured at high redshifts with observables which are relatively easy to obtain (albeit the high calibre data of observables is vital), we develop a environment-dependent selection criteria to obtain a statistically large sample of BCG progenitors beyond $z \sim 1$.


Our basic idea is to select the densest environments at high redshifts and identify our BCG progenitors as the most massive galaxies in these most overdense  environments. For a complete observational galaxy sample at high redshifts, environmental density can be measured for each galaxy through galaxy counts within a fixed physical aperture.  Once the densest environments are located, we select the most massive galaxy in each cylinder as the BCG progenitor candidate. The summary of this method is that once the environmental densities for all galaxies are obtained, they are ranked from the largest overdensity down to the smallest overdensity. Given the volume of the CANDELS survey and using the number densities of galaxy clusters in the local universe, there will be a number of $N$  BCG progenitors that need to be selected as the most massive galaxies in the top $N$ densest regions.

We apply our method to the observational data of the CANDELS UDS. At a constant number density of $10^{-4.06} h^3$Mpc$^{-3}$, 38 progenitors need to be selected at $z=1-3$, as we discussed in Section~\ref{sec:constant_ND}. The environmental densities have already been measured by \citet{Lani13} for the UDS which covers the CANDELS UDS. Therefore, the densities are known for CANDELS UDS galaxies. By ranking them from the most overdense to the least overdense, we select  $38$ progenitors as the most massive galaxies in the top $38$ densest regions. We then compare this sample with the $469$ local descendants from our SDSS DR7 sample.   


 
Although we can obtain a progenitor sample this way, it is possible that a fraction of these galaxies are not the true BCG progenitors but are the progenitors of non-BCGs at $z\sim 0$. Important questions are  how many true BCG progenitors are in our selected high-$z$ samples and what fraction of the true BCG progenitors are selected. We carry out a series of tests in the  simulations to answer these questions.

\subsection{Test of Method in Simulations}
\label{sec:test_in_simulation}
To test our assumption of the BCG progenitor selection, we use the output of the Millennium Simulation and their respective SAM realisations. The Millennium Simulation uses $2160^3$ particles of mass $8.6 \times 10^8 h^{-1} M_\odot$ to follow the evolution of the DM distribution within a comoving box of side $500 h^{-1}$Mpc from $z=127$ to $z=0$ in $64$ snapshots. Using the assumption of the $\Lambda$CDM cosmological model, the cosmological parameters are $\Omega_{\rm m}=0.25$, $\Omega_{\rm b}=0.045$, $\Omega_{\rm \Lambda}=0.75$, $h=0.73$, $\sigma_8=0.9$ and $n_{\rm s}=1$. The SAM used in this work is from \citet{DeLB07}. They study the formation and evolution of BCGs by applying their model to the output of the Millennium Simulation with the updated treatments for stellar populations, dust attenuation and cooling flow suppression via AGN feedback. 

We employ the simulation data at two redshift snapshots. One is $z=0$ (snapshot=63) at which we identify a sample of BCGs. All of their progenitors can be traced easily at any higher redshift. The other epoch we study is  $z=2.07$ (snapshot=32) at which we select the progenitor sample by using the same method that we use on our data. Although the $38$ progenitors from the CANDELS UDS are chosen from $z=1- 3$,  their average redshift is $z=2.06$ (see Fig.~\ref{fig:candelsz}). Thus the simulation comparison is carried out at the SAM snapshot at $z=2.07$. In the following, we describe in detail how we define the galaxy sample used in the tests at $z=0$ and $z=2.07$. We then discuss the fraction of true BCG progenitors which are selected by our method. We also examine and discuss the fraction of BCGs recovered when densities measured with different parameters  are used, or when the top three most massive galaxies are identified as the BCG progenitor candidates. The implication of these results will be discussed briefly.

\subsubsection{Simulation Snapshot at  z=0 Sample Selection}
\label{sec:snapshot_z0}
In the full simulation box at $z=0$  BCGs are identified as the most massive galaxy within the virial radius of their DM halos whose mass $M_{\rm vir} \geq 10^{13.55} h^{-1} M_\odot$. This halo mass criteria is employed to be consistent with the observational halo mass which is $M_{200} \geq 10^{13.55} h^{-1} M_\odot$ corresponding to $10^{-4.06} h^3$Mpc$^{-3}$ (see Section~\ref{sec:constant_ND}). There are $8490$ BCGs identified at $z=0$ in the simulation through this method. 

Once the BCGs at $z=0$ are selected, it is straightforward to trace their progenitors at any higher redshift. For the full comparison between observations and simulations, we use the observational constrains in the simulations. There is a stellar mass cut of $M_*^{\rm cut}=10^{9.76} M_\odot $ at $z\sim 2$ for the galaxy completeness used in \citet{Lani13} (see Section~\ref{sec:density_obs}). To be consistent with this, at $z=2.07$ in the simulation, we only consider galaxies whose stellar mass $M_* \geqslant 10^{9.76} M_\odot $ to ensures that the parent high-$z$ galaxies used in the simulation to calculate the environmental density is as similar to the observational one as possible. Then every galaxy at $z=2.07$ which ends up as one of our local $8490$ BCGs is counted as a true BCG progenitor. At $z=2.07$ there are $78,454$ true BCG progenitors in total for the whole $8490$ BCGs, comprising a ``true BCG progenitor catalogue''.  These progenitors however are not only the most massive progenitors, but include all the individual objects that grow and merge to form the BCGs in the local universe within the simulation.  

\subsubsection{Snapshot of z=2.07}
\label{sec:snapshot_z2}

\begin{figure*}
\center{\includegraphics[scale=0.22]{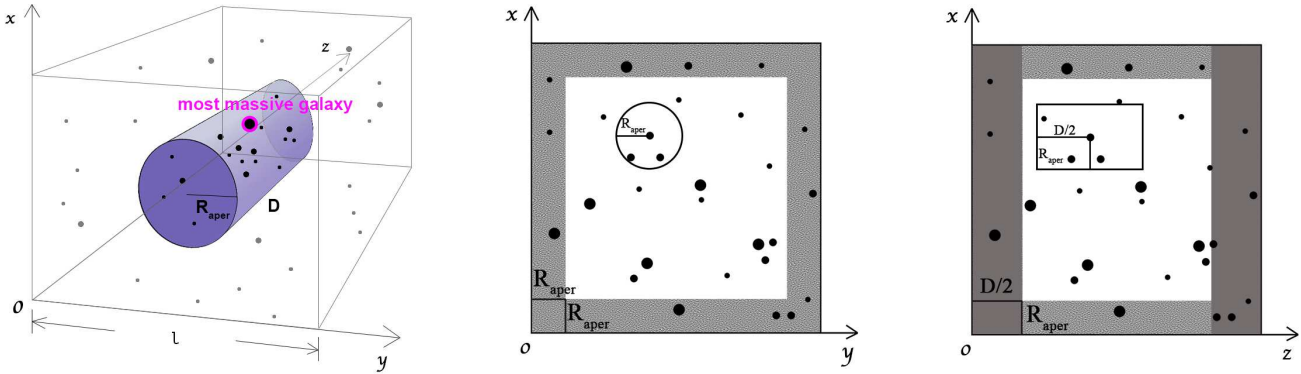}}
\caption{Left panel illustrates how density is measured through galaxy counts (black dots) in a cylinder of fixed aperture $R_{\rm aper}$ and depth $D$ (coloured in purple) for the central galaxy. Larger dots show the more massive galaxies. The most massive galaxy in the cylinder, with a magenta circle, is selected as the BCG progenitor candidates in this example. The density is measured in the z-direction of the box in the simulation. The grey dots are the galaxies outside the cylinder. Middle panel shows a cross section perpendicular to the z-axis. Density is not calculated for those galaxies in the shaded area whose vertical distance to the box edges of the x-axis or y-axis is less than the aperture radius $R_{\rm aper}$. Right panel shows a cross section perpendicular to the y-axis. Density is not calculated for those galaxies in the dark shaded area whose distance in z direction to the x-y surface is less than $D/2$. In all, galaxies in the dark and light shaded area in the right panel are excluded from the density catalogue in the simulation. The density of galaxies in the inner white region is measured within a fixed aperture, as shown in the left panel.}
\label{fig:densitycubic}
\end{figure*}

In order to apply our observational method on the simulation data to select BCG progenitor candidates, we first calculate the environmental density for galaxies in the full simulation box at $z=2.07$ by using galaxy counts in a fixed physical aperture, with some modification of  Equation~\ref{eq:density_UDS}. In the simulations, there are no bad pixels that need to be masked as in the observations. Thus the $N_{\rm mask}^{\rm aper}$ term is the area of the chosen aperture, and  $N_{\rm mask}^{\rm tot}$ is the area of one side of the full box. The term  $N_{\rm mask}^{\rm tot}/ N_{\rm mask}^{\rm aper}$ then reduces to $l^2/ (\pi R_{\rm aper}^2)$, where $l$ is the box length of one side (i.e., 500 $h^{-1}$Mpc) and $R_{\rm aper}$ is the aperture radius. We use a density contrast in our test defined as

\begin{equation}
\delta = \frac{N_{\rm g}^{\rm aper}}{N_{\rm z}} \times \frac{l^2}{\pi R_{\rm aper}^2} - 1 ,
\end{equation} 

\noindent where $N_{\rm g}^{\rm aper}$ is, as before, the number of galaxies in the chosen aperture. $N_{\rm z}$ is the total number of galaxies within $l^2 \times D$, following the definition in \citet{Lani13}, where $D$ is the depth of the cylinder. The cylinder we use is in the direction of the z-axis. The left panel of Fig.~\ref{fig:densitycubic} illustrates how the density is measured within a fixed aperture. The most massive galaxy in the cylinder (circled in magenta) is a  BCG progenitor candidate as we  discuss in Section~\ref{sec:basic_idea}. 

The values of the aperture radius $R_{\rm aper}$, and the depth of the cylinder $D$, are chosen to be similar to the ones adopted  in \citet{Lani13} who construct a cylinder with an aperture radius of $400$ kpc, and depth of $1$ Gyr to measure  UDS densities. In the simulation,  the value of $R_{\rm aper}=400$ kpc can be employed easily. However, it is difficult to apply a $1$ Gyr depth as the cylinder depth in one single box. Unfortunately,  the $1\sigma$ uncertainty of the UDS photometric redshifts $\Delta z \sim 0.1$ at $z\sim 2$ corresponds to $\pm 300 h^{-1}$Mpc ($\Delta v \sim \pm 30000$ km/s).  This is already larger than the box size in each redshift snapshot.   The depth of 1 Gyr is thus several times greater than the $1\sigma$ uncertainty of the UDS photometric redshift. Since we are limited by the simulation box, we are more generous in considering the photometric redshift errors at high redshift. Additionally, to ensure a large sample of galaxies in the simulation box being eligible to have a reliable density measurement, we use $D=120 h^{-1}$Mpc as the cylinder depth.

As we mention in Section~\ref{sec:snapshot_z0}, only galaxies whose mass $M_* \geqslant 10^{9.76} M_\odot $ will be considered within the $z=2.07$ selection in the simulation. However, density is not measured for galaxies too close to the box edges, as a full measure of environment cannot be done.  Therefore, there is no measurement of environmental density for  galaxies whose perpendicular distance to the box edges in the x-axis or y-axis is less than the chosen aperture radius. The middle panel of Fig.~\ref{fig:densitycubic} shows a cross section perpendicular to the z-axis. The galaxies in the shaded area are excluded from the density catalogue. 


On the other hand, if the distance in the z direction from one galaxy to the x-y surface is less than $D/2$, the density measurement will not be employed on this galaxy for the same reason that no galaxy information can be traced in the space outside the simulation box. A cross section perpendicular to the y-axis in the right panel of Fig.~\ref{fig:densitycubic} illustrates this requirement on distance in the z-direction. Finally, a density catalogue which is ranked from the  largest densities to the smallest densities is created for galaxies with $M_* \geqslant 10^{9.76} M_\odot $ at $z=2.07$ in the simulation. The most massive galaxy is known in each density and is taken as the BCG progenitor candidate in our selection.  

In the simulation, environment number density tracing is applied as well. Since there are $8490$ BCGs/clusters at $z=0$ in the simulation, we need to select $8490$ environments/progenitors within the $z=2.07$ snapshot. Based on our assumption for BCG progenitors, the progenitor sample galaxies are identified as the most massive galaxies in the top $8490$ densest environments in the simulation. Finally, the number of true BCG progenitors within our observationally based selection sample can be known by matching our $8490$ progenitors  with the $78,454$ true BCG progenitors the simulation gives us.  

\subsubsection{Fraction of the Selected True BCG Progenitors}
\label{sec:prog_fraction}

In order to show explicitly how many true BCG progenitors can be found with our observationally based selection method, we define

\begin{equation}
f_{\rm tot} = N_{\rm tot}^{\rm match} / N_{\rm tot},
\end{equation} 

\noindent which is a fraction of the true BCG progenitors in our selected high-$z$ progenitors. The term $N_{\rm tot}=8490$ is the total number of the progenitor galaxy sample identified through our method of environmental matching on the most massive galaxies in the densest environments. $N_{\rm tot}^{\rm match}$ is the number of true BCG progenitors found within these $N_{\rm tot}=8490$ progenitors. Note that we allow more than one BCG progenitors to end up in the same local BCG. 

In the end, we find $f_{\rm tot} = 45$\%, indicating that the progenitor sample selected by our density-dependent method at $z \sim 2$ is not a pure sample of  true BCG progenitors, but is contaminated by $55$\% of progenitors of local non-BCG galaxies (we call these systems non-BCG progenitors hereafter).  If we select the high-$z$ progenitor sample by the fixed galaxy number density (the method when study massive galaxy evolution), the fraction of true BCG progenitors is only $35$\%. Our method can obtain a much higher fraction of successful BCG progenitor selection. In Section~\ref{sec:nocontam_highz}, we will examine how different the properties are between our true BCG progenitors and the non-BCG progenitors, as well as discuss whether the progenitor sample we select can be used to trace BCG evolution and how to account for this contamination when comparing high and low redshifts.


\subsubsection{Effect of Density Measurement Method} 
\label{sec:effectofdensity}

The $f_{\rm tot}$ value we measure is based on the density measurement with a cylinder size of $R_{\rm aper}=400$ kpc and $D=120 h^{-1}$Mpc. In this section, we examine whether the cylinder size can significantly effect the fraction of  the selected true BCG progenitors using our method. We thus apply different apertures and depths to calculate the environmental density.

First, with a fixed depth of $D=120 h^{-1}$Mpc, we employ different aperture radii  from the local scale $R_{\rm aper} = 250 $ kpc, to the global scale $R_{\rm aper} = 1$ and $2 $ Mpc. The total fraction of selected true BCG progenitors is then $f_{\rm tot}= 47$\%, $42$\% and $39$\%, respectively for these different scenarios.  It thus appears that $f_{\rm tot}$ increases at smaller aperture radii, however the aperture size is not a major factor in significantly increasing the number of selected true BCG progenitors.

We also measure galaxy number densities within cylinders with a fixed aperture of $R_{\rm aper} = 400$ kpc and with different depths of $D=250$, $30$, and $4 h^{-1}$Mpc. $250h^{-1}$Mpc represents the largest photometric redshift uncertainty which we have in the data. $30h^{-1}$Mpc is of the same order of  redshift accuracy measured by narrow-band imaging, and $4 h^{-1}$Mpc is the spectroscopic redshift measuring error. The corresponding total fractions are then: $f_{\rm tot}= 36$\%, $51$\% and $56$\%, respectively. This implies that if spectroscopic redshifts for a large sample of galaxies in the early universe could be measured  accurately in observations, the fraction of selected true BCG progenitors could increase by $> 10$\% compared to using SED-fitted photometric redshifts. However, the fraction of  true BCG progenitors selected  as the most massive galaxies in the densest environments cannot exceed $70$\% even if we use a cylinder with very small aperture (e.g., $R_{\rm aper} = 250$ kpc) and a spectroscopic redshift uncertainty (e.g., $4 h^{-1}$Mpc).   This suggests that there is a natural limit in how well we can trace BCG progenitors with this method.

The length of the cylinder we use to measure density in the observations is equivalent to 1 Gyr of look-back-time (or $\sim 1000h^{-1}\,$Mpc around $z\sim 2$), which is significantly larger than the one we have used in the simulations due to the size of the simulation box. In order to use a cylinder with a more similar length to the one used in the observations, we have replicated the $z=2.07$ simulation box on all sides, taking advantage of the periodic boundary conditions. This allows us to measure environmental density with cylinder whose length is more than $500h^{-1}\,$Mpc. By using cylinder lengths of $500h^{-1}\,$Mpc and $1000h^{-1}\,$Mpc, the fraction of the true BCG progenitors in our selected high-$z$ sample is $\sim41$\% in both cases, a number that is very similar to what was found with smaller cylinders. We are thus reassured that our results on BCG evolution do not depend on the exact size of the cylinder used in the simulation tests.

\subsubsection{Effect of Galaxy Stellar Mass}
\citet{DeLB07} show in simulations that BCG progenitors have a wide stellar mass distribution from $10^{10} M_\odot$ to $10^{12} M_\odot$, and there is a good overlap between the mass distribution of high-$z$ massive galaxies and the massive progenitors of local BCGs. This implies that the most massive galaxy in a dense region could  be a non-BCG progenitor and we could miss out those true BCG progenitors whose stellar masses are slightly smaller. 

Therefore, in the simulation, we select the candidates of BCG progenitors from a larger pool that includes the second and third most massive galaxies in the densest regions to examine the possible effect from stellar mass differentials. We carry out this test through a method of iterative matching.  We first test if the most massive galaxy in a given environment is a BCG progenitor.  If this most massive galaxy is matched as the true BCG progenitor then we do not further match  the 2nd and 3rd massive galaxies. However, if the top massive galaxy is a not a BCG progenitor then we match the 2nd most massive galaxy in that environment with the true BCG progenitors. No further matching will be done on the 3rd galaxy as long as the 2nd one is the true BCG progenitor. If neither the 1st or 2nd massive galaxies are  BCG progenitors then we match  the 3rd most massive one. This selection down to the 3rd most massive galaxy increases the total fraction of the true BCG progenitors we select to $f_{\rm tot} = 55$\%.

Combined with the results of Section~\ref{sec:effectofdensity} this indicates that a large fraction of massive galaxies in very dense environments at high redshift do not end up in $z=0$ BCGs but in local normal massive galaxies. Both  overdensity and  stellar mass are not unique tracers for  identifying true BCG progenitors at $z \sim 2$. Other than using environmental density, we also examine the fraction of true BCG progenitors in the simulation if our progenitors are selected based on their host DM subhalo masses.  \citet{Muldrew11} demonstrate that in simulations the maximum circular velocity of the subhalo is a better property to represent the subhalo mass than the virial mass of subhalo.  We thus examine the selection that the BCG progenitors are selected as the $8490$ most massive galaxies in the top $8490$ subhalos sorted by their maximum circular velocity. If we use this method, the total fraction of the selected true BCG progenitors increases  to $f_{\rm tot} = 65$\%. Although dark matter is a more promising tracer to find BCG progenitors at $z \sim 2$, it is hard to apply it on observation data since measuring the maximum circular velocity of subhalo cannot be done observationally at the moment. However, ultimately this may be a better method of finding BCG progenitors in the future. 
 

\subsection{Effect of Contaminants in Our Selected Sample}
\label{sec:contam_effect}

Since our final aim is to apply our BCG progenitor selection on the CANDELS UDS data by employing the UDS density catalogue, the following discussion will be based on the results of the simulation tests in Section~\ref{sec:prog_fraction}. We show that using the density measured as in \citet{Lani13} within the UDS, our selected progenitors at $z \sim 2$ are not pure BCG progenitors, but consist of $45$\% true BCG progenitors and $55$\% non-BCG progenitors as contaminants. This means that within the $38$ progenitors selected from the CANDELS UDS at $1 \leqslant z \leqslant 3$, about $17$ of them are BCG progenitors and the rest are contaminants.  It is, however, impossible to know from the available data which are the real BCG progenitors and which are not.  

In the following we first show, based on simulations, that the properties of our entire selected progenitor sample and the  $45$\% true BCG progenitors within them  are not significantly different.  Next, in order to trace BCG evolution down to $z \sim 0$, our selected progenitors at high redshifts need to be compared with their counterparts in local universe, which will be a mixture of BCGs and non-BCGs. We demonstrate below that the local non-BCGs which are the descendants of those $55$\% non-BCG progenitors statistically share similar properties of local BCGs. We find that the uncertainty resulting from the contamination in our samples does not erase the BCG evolution signal. The comparison, at the same number density, between the progenitors we select at high redshift with their local counterparts can therefore give us an accurate measurement of BCG evolution.


\subsubsection{Contamination at High-$z$}
\label{sec:nocontam_highz}

\begin{figure*}
\center{\includegraphics[scale=0.45]{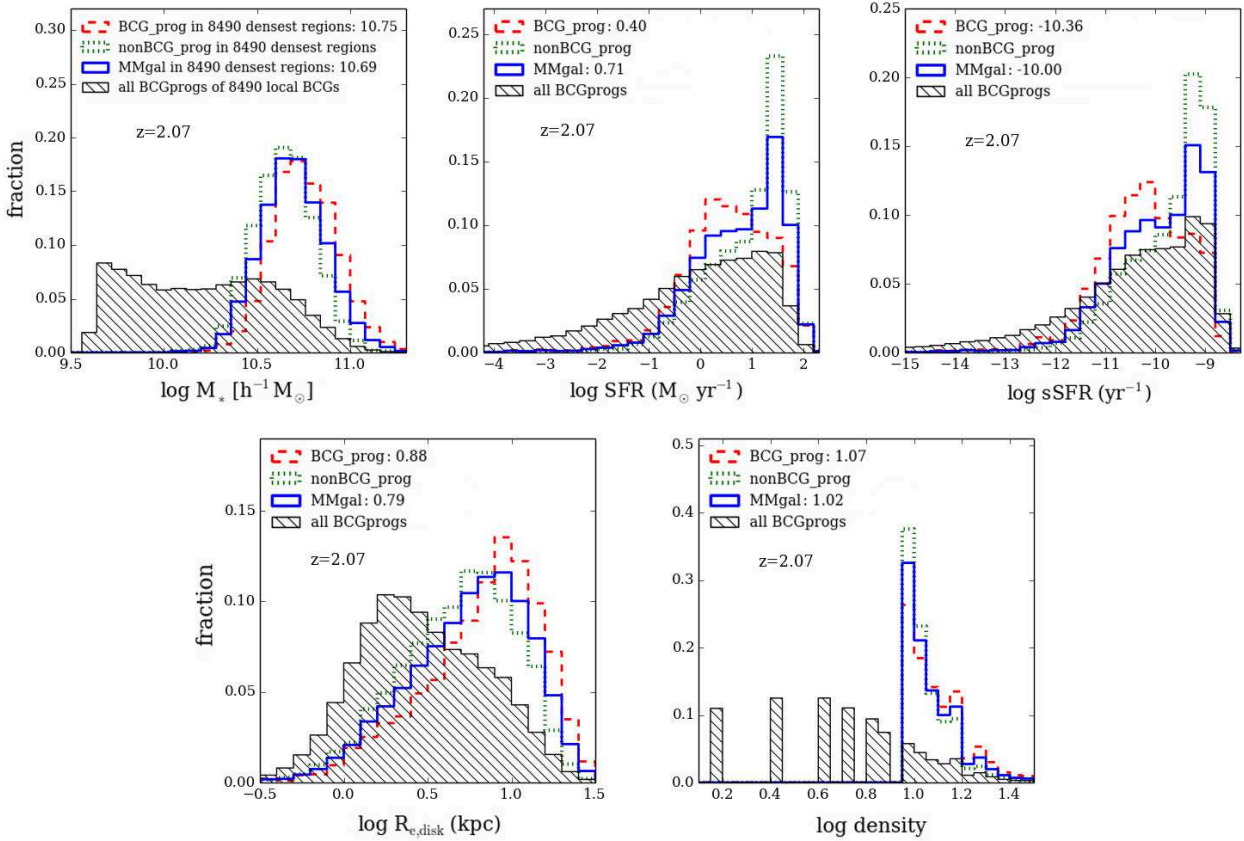}}
\caption{Property distributions of the 8490 most massive galaxies in the top 8490 densest regions (blue solid) at $z=2.07$ from the Millennium simulation. In the upper-row, stellar mass, SFR, and sSFR are shown in panels from left to right, respectively. In the lower-row, the left panel is the disk radius distribution, and the right panel is for density measured within a fixed aperture. In each panel,  distribution of the 45\% true BCG progenitors within these 8490 galaxies is illustrated as the red dash line. The remaining 55\% non-BCGs progenitors are presented in the green dotted line. The black line with shadow presents the distribution of the entire 78,454 progenitors of local 8490 BCGs in the simulation. The numbers in legend show the median value of the corresponding distributions.  Detail discussions can be found in text.}
\label{fig:similarhighz}
\end{figure*}

From the test we carry out in Section~\ref{sec:test_in_simulation}, we know that the $38$ progenitors that we select from the CANDELS UDS by our method is a mixed sample with 45\% true BCG progenitors and 55\% non-BCG progenitors. The question we need to answer is how the contaminant non-BCG progenitors differ from the true BCG progenitors. Ideally, the comparison should be carried out in observational data between the whole $38$ progenitors and the $17$ true BCG progenitors within them. However, there is no method that can identify the true BCG progenitors in our selected high-$z$ sample. Therefore, we carry out our comparison in simulation by using the $8490$ progenitors selected through the observationally based selection. They are compared with the $3780$ (45\%) true BCG progenitors, and $4710$ (55\%) non-BCG progenitors within them (see Section~\ref{sec:snapshot_z2} and Section~\ref{sec:prog_fraction}). The properties within the simulation we discuss are stellar mass, SFR/sSFR, disk radius, and density and position of galaxies. 


First, we examine the differences in galaxy masses.  The left panel in the upper-row of Fig.~\ref{fig:similarhighz} illustrates the stellar mass distribution of our selected $3780$ true BCG progenitors (red dash line) and the $4710$ non-BCG progenitors (green dotted line) in the simulation. It illustrates that the true BCG progenitors selected by our method are slightly more massive than those selected which are non-BCG progenitors. The non-BCG progenitors make the entire $8490$ progenitors (shown blue solid line) have on average a somewhat smaller stellar mass.  The median stellar mass of true BCG progenitors is $10^{10.75} h^{-1}$ M$_\odot$, and it is $10^{10.69} h^{-1}$ M$_\odot$ for all the $8490$ progenitors. The effect of non-BCG progenitors on the stellar mass distribution is thus to make it $0.06$ dex smaller. Moreover, we also plot the mass distribution of the entire $z=2.07$ progenitor population of the 8490 $z=0$ BCGs (i.e., the 78,454 true BCG progenitors. See Section~\ref{sec:snapshot_z0}). This is shown in Fig.~\ref{fig:similarhighz} as the black shaded area. It is clear that our method selects those BCG progenitors at the most massive end.     

The next two properties we examine are SFR and sSFR, whose distributions are shown in the middle and right panels in the upper-row of Fig.~\ref{fig:similarhighz}. As can be seen, the non-BCG progenitors, which make up $55$\% of the selected sample (green dotted line), and the entire selected samples (blue solid line) have a different distribution from the $45$\% true BCG progenitors (red dash line). The actual BCG progenitors distribute relatively evenly over $\log SFR= [0,2]$ ($\log sSFR = [-11,-9]$), with a larger fraction found towards the low SFR and low sSFR values. If we take the median SFR of the $8490$ progenitors as a threshold, the majority of the true BCG progenitors have a SFR lower than $\log SFR = 0.71$ and a similar fraction for sSFR selection. In contrast, the selected non-BCG progenitors and the whole progenitor sample are dominated by galaxies with high SFR (high sSFR). The non-BCG progenitor population makes the SFR (sSFR) distribution of the entire selected progenitors larger by a factor of $\sim 0.3$ dex ($\sim0.4$ dex) than the true BCG progenitors. Although the exact difference in SFR/sSFR values between true BCG progenitors and our  observationally-selected BCG progenitors is likely  dependent on the SAM used, it is unlikely that the uncertainty introduced by this is responsible for the clear SFR evolution that we detect in Section~\ref{sec:results} (by almost two orders of magnitude).

The left panel in the lower-row of Fig.~\ref{fig:similarhighz} shows the distribution of disk radius which is derived by \citet{DeLB07} from halo radius following the relationship in \citet{Mo98}. We find that non-BCG progenitors (green dotted line) in our selected sample tend to have smaller disk radii, making the entire sample of selected progenitors (blue solid line) more compact in disk size by a factor of $0.1$ dex than the true BCG progenitors (red dash line). At the same time, it is clear that the true BCG progenitors we select from the densest environments have much larger radii compared with the entire 78,454 true BCG progenitors of the 8490 $z = 0$ BCGs (black shaded area).


Moreover, we also check the environments of our selected samples in the simulation. The distribution of density where our selected progenitors reside is presented in the right panel in the lower-row of Fig.~\ref{fig:similarhighz}. We find that the environments of the $45$\% true BCG progenitors (red dash line) are only marginally denser than the environments which host the $55$\% non-BCG progenitors (green dotted line). About 10\% more non-BCG progenitors are in the less dense regions. Nevertheless, a K-S test demonstrates that the entire sample of our selected progenitors (blue solid line) are within  the same local density as the 45\% true BCG progenitors.   This result is partially by design given that we only select our progenitors based on being in dense environments. However it might be the case that the BCG progenitors are more likely to be found in the densest environments among this selection, but this appears to not be the case as presented by the black shaded area which shows environments of the entire 78,454 true BCG progenitors of the 8490 $z = 0$ BCGs. It is clear that majority of the true BCG progenitors reside in less dense regions.

In addition to the environmental density, the location of galaxy in the host dark matter halo is examined  as well. The simulation gives the central galaxy of its FOF group as type 0, the central galaxy of a subhalo is type 1, and satellite galaxy as type 2. In the $3780$ true BCG progenitors we select, 71\% of them are type 0 galaxies while 19\% are type 1 galaxies and the rest 10\% are type 2 galaxies. In the $4710$ non-BCG progenitors, we find that 72\% are type 0 galaxies, 23\% are type 1 galaxies and other 5\% are type 2 galaxies. There is thus not much difference in terms of galaxy position within their respective groups and clusters between BCG progenitors and non-BCG progenitors.

Based on the simulation, we find that the properties of the entire progenitor population selected by our method are very similar to the properties of the actual BCG progenitors within them.  These properties include: stellar mass, disk radius and environment. The non-BCG progenitors do however appear to influence the distribution of SFR/sSFR, driving the SFR/sSFR of the entire selected progenitors higher by a factor of $0.3 - 0.4$ dex larger. 

We apply these findings on our $38$ observational progenitors, supposing that their stellar masses and effective radii represent the true BCG progenitor at $z\sim 2$ but with a $\sim 0.4$ dex larger SFR/sSFR.   In Section~\ref{sec:BCGevo_intrinsic} we demonstrate that the evolution of  BCGs over $z= 0 - 2$ is intrinsic, and still evident, even if the systematic raising of the star formation from the non-BCG progenitors is considered.

\subsubsection{Contamination in the Local Universe}
\label{sec:nocontam_local}

Since the progenitor population selected by our method is a mixture of real BCG progenitors and non-BCG progenitors, in order to trace evolution down to $z\sim 0$, the local comparison should be the $z \sim 0$ counterparts of our high-$z$ progenitors rather than a pure local BCG sample.  In this section, we discuss how we construct an observational local mixed sample  which consists of the descendants of both our selected high-$z$ non-BCG progenitors and the BCG progenitors.  We also examine whether the properties of a locally mixed sample are different from a pure BCG sample due to the non-BCG contamination. At our constant number density selection of $10^{-4.06} h^3$Mpc$^{-3}$, we calculate that $469$ local descendants should be selected from SDSS DR7.   We must populate these descendants with both real BCGs and with other non-BCG galaxies.  

In the simulation, we find that within the $z=0$ descendants of the $z=2.07$ 8490 selected progenitors found using our observational method, $38$\% of them are BCGs and the remaining $62$\% are non-BCGs. Applying these fractions  to the 469 local observational sample, there are thus 291 non-BCGs and 178 BCGs we should identify to build up a mixed counterpart sample at $z \sim 0$.

In order to ensure that the $291$ local non-BCGs we identify in SDSS DR7 catalogue are likely the descendants of our selected high-$z$ contaminants from CANDELS UDS, they are chosen according to their distribution within the whole $z=0$ galaxy population in terms of stellar mass. This distribution can be determined based on our simulation results. In the simulation, at $z=2.07$, 4710 galaxies ($55$\%) selected by our method in the top 8490 densest regions evolve into non-BCGs at $z=0$. In terms of stellar mass, how these non-BCGs distribute in the whole $z=0$ galaxies can be known. By ranking  galaxies by stellar mass from large to small in the $z=0$ box, descendants of our non-BCG progenitors are located by their  stellar masses (hereafter we call the mass-ranked whole local galaxy population as the ``galaxy pool''). 

\begin{figure*}
\centering{ 
\center{\includegraphics[scale=0.46]{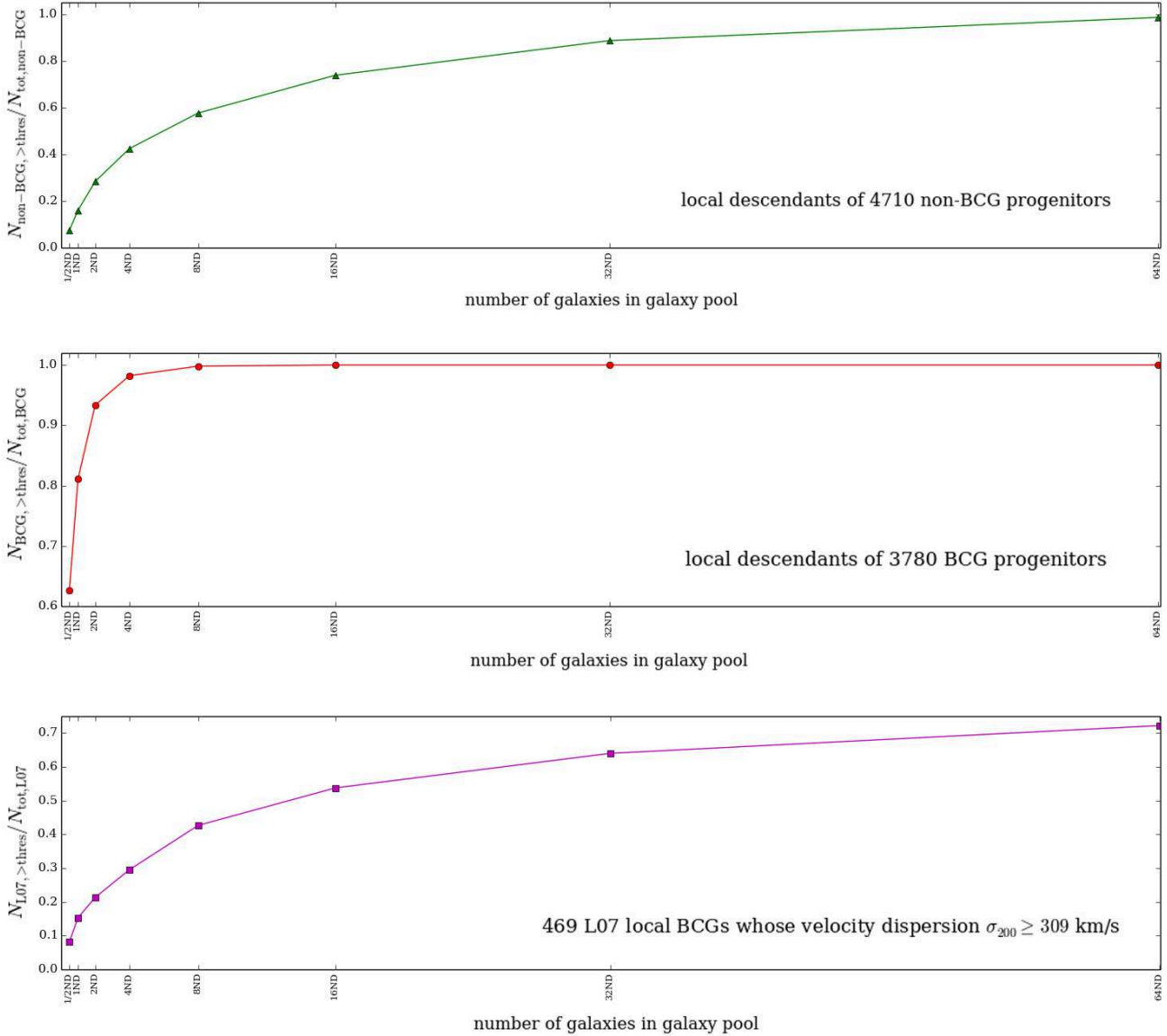}}
\caption{\textit{Upper} panel: cumulative fraction of local non-BCGs which are the descendants of the 4710 $z=2.07$ non-BCG progenitors in the simulation.  These galaxies are plotted as a function of galaxy number in the ``galaxy pool'' (see text) from our simulation. Galaxies in the galaxy pool are ranked by their stellar masses, from large to small. Note that the galaxy number in the galaxy pool can easily be converted to a number density by dividing the volume of the $z=0$ simulation box ($500^3 h^{-3}$Mpc$^3$). At the number density of $ND=10^{-4.06} h^{3}$Mpc$^{-3}$, we choose the numbers, such that they can be converted as  $m \times ND$ where $m=1/2, 1, 2, 4, 8, 16, 32, 64 $. The tick label of the x-axis is thus expressed in terms of  $m \times ND$.  Searching from the most massive galaxy down to a chosen number of galaxies (i.e., $m \times ND$), the number of descendants we retrieve from our selected non-BCG progenitors is obtained, which is expressed as $N_{\rm non-BCG, >thres}$. This number $N_{\rm non-BCG, >thres}$ can be converted into a cumulative fraction defined as $f_{\rm non-BCG, >thres} = N_{\rm non-BCG, >thres} / N_{\rm tot, non-BCG}$ where $ N_{\rm tot, non-BCG}$ is the total number of $z=0$ descendants of the 4710 non-BCG progenitors.  When the y-axis reaches $= 1$ this is when all of the descendants of non-BCG progenitors have been recovered. \textit{Middle} panel: cumulative fraction of simulated $z=0$ BCGs which are the descendants of the 3780 $z=2.07$ BCG progenitors as a function of galaxy number density in the simulated galaxy pool. \textit{Lower} panel: real data for BCGs, showing the cumulative fraction of local L07 BCGs as a function of galaxy number in the SDSS DR7 galaxy pool.}
\label{fig:findcontam_mass_distrb_L07BCG}}
\end{figure*}

Therefore, down to a specific stellar mass $M_{*,\rm threshold}$ in the galaxy pool, we could know how many non-BCGs whose $M_* \geqslant M_{*,\rm threshold}$ are there. Note that, in the galaxy pool, a specific stellar mass $M_{*,\rm threshold}$ corresponds to a number of galaxies whose $M_* \geqslant M_{*,\rm threshold}$. Taken into account the local comoving volume,  a specific stellar mass $M_{*,\rm threshold}$ then corresponds to a cumulative number density $ND(>M_{*,\rm threshold})$. Since we use the number density of  $ND=10^{-4.06} h^3$Mpc$^{-3}$ in this work, to be convenient, we take this value as a unit. When we explore the mass distribution of non-BCGs in the galaxy pool, we choose a number of $M_{*,\rm threshold}$ whose converted cumulative number densities are $m \times ND$ where $m=1/2, 1, 2, 4, 8, 16, 32, 64 $. In the upper panel of Fig.~\ref{fig:findcontam_mass_distrb_L07BCG},  the lower tick labels of x-axis show these corresponding cumulative number density.

Down to each $M_{*,\rm threshold}$, the number of descendants of our selected non-BCG progenitors can be obtained (we express this as $N_{\rm non-BCG, >thres}$). This number can be converted to a cumulative fraction defined as $f_{\rm non-BCG, >thres} = N_{\rm non-BCG, >thres} / N_{\rm tot, non-BCG}$ where $ N_{\rm tot, non-BCG}$ is the total number of $z=0$ descendants of our $4754$ non-BCG progenitors.   This fraction is the y-axis of the upper panel of Fig.~\ref{fig:findcontam_mass_distrb_L07BCG}.  The upper panel of Fig.~\ref{fig:findcontam_mass_distrb_L07BCG} finally shows the distribution of the $z=0$ descendants of our selected non-BCG progenitors in  the galaxy pool in terms of stellar mass.  If the total number of non-BCGs is known (i.e., $N_{\rm tot, non-BCG}$ is known), this panel essentially tells us how many non-BCGs are between two adjacent cumulative number densities of galaxy pool. This panel also tells us we need to go down to $64ND$ in the galaxy pool to retrieved almost all the descendants of our non-BCG progenitors.

We apply this distribution to the SDSS DR7 galaxies to select the observational descendants of our high-$z$ non-BCG progenitors. In observation, the galaxy pool is comprised of the  SDSS DR7 galaxies which are at  $0.02 \leqslant z \leqslant 0.1$ and are ranked by  stellar mass from large to small. The cumulative number densities $m \times ND$ where $m=1/2, 1, 2, 4, 8, 16, 32, 64 $ are also used for the SDSS DR7 data. For each cumulative number density, the corresponding number of SDSS DR7 galaxies (counted from the most massive galaxy) is known, from which the non-BCGs are selected. Since we need to obtain $291$ non-BCGs to contaminate our pure BCGs (i.e., $N_{\rm tot, non-BCG}=291$), how many non-BCGs should be selected between two adjacent cumulative number densities could be known according to the upper panel of Fig.~\ref{fig:findcontam_mass_distrb_L07BCG}. The  non-BCGs are then selected randomly from galaxies which are not BCGs.



There is also the caveat that the galaxy distribution in our simulation cannot fully represent the observational one due to the unclear baryon physics in galaxy formation and evolution. However, the distribution of local descendants of high-$z$ non-BCG progenitors from this simulation is currently the best method we can take for selecting non-BCG descendants in observations. Moreover, since the formation and evolution of non-BCGs may involve less hydrodynamical mechanisms such as inflows/outflows at $z < 3$, the simulation results for non-BCGs could be better than the results for BCGs.

The question now is how do we select the BCG themselves at low redshifts?  The middle panel of Fig.~\ref{fig:findcontam_mass_distrb_L07BCG} shows the simulated cumulative number distribution of $z=0$ BCGs which are the descendants of our selected $3780$ $z=2.07$ BCG progenitors in the simulation. However, this simulation and others create too many massive galaxies compared with observations at $z\sim 0$ (e.g., \citealt{Lin13}), such that the simulation distributions of local massive galaxies does not represent the real observational ones correctly.  This is shown clearly by comparing the middle panel of Fig.~\ref{fig:findcontam_mass_distrb_L07BCG} with the lower panel which illustrates the L07 BCG number distribution in the SDSS DR7 galaxy population. Therefore, we select 178 BCGs from the L07 catalogue according to the L07 BCG number distribution.  The BCGs within every bin are selected randomly from those galaxies which are L07 BCGs whose host clusters have velocity dispersion $\sigma_{200} \geq 309$ km/s.

\begin{figure*}
\centering{ 
\subfigure{
\raggedright{\includegraphics[scale=0.59]{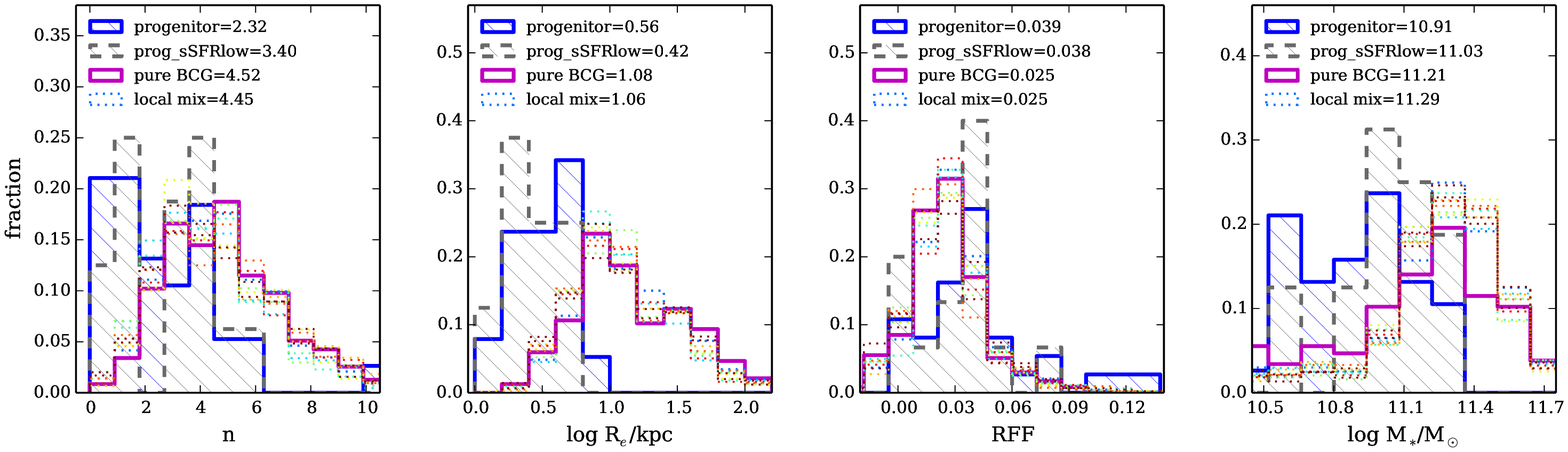}}}  
\subfigure{
\centering{\includegraphics[scale=0.58]{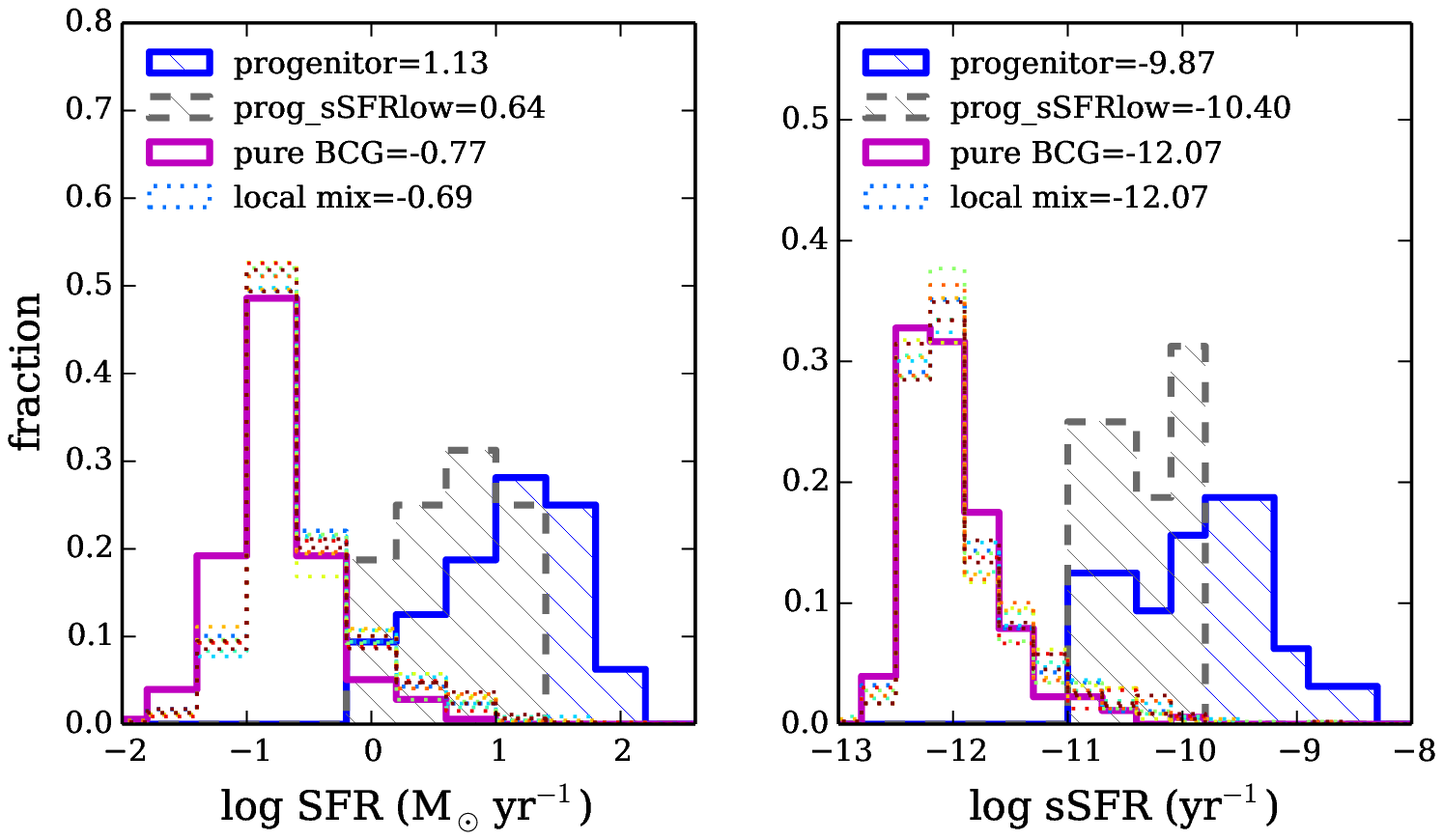}}}
\caption{Distribution of \sersic\ index $n$, effective radius, \rff , stellar mass, SFR and sSFR from our observations. The blue solid line with shadow in each panel shows the property distribution of the 38 high-$z$ progenitors selected by our method from CANDELS UDS. Specifically, the grey dashed lines with shadow illustrate the distributions of our selected progenitors whose sSFR is lower than the median value (i.e., $\log sSFR < -9.87$). The several colour dotted lines show the distributions of the 10 sets of local mixed sample each of which contains 291 non-BCGs and 178 BCGs. Magenta solid line is for the 178 pure BCGs in one set of the local mixed sample. The legends indicate the median value of each distribution. The value for the local mixed sample in the legend is the average median value of the 10 sets. Detailed discussions  are in the text.}
\label{fig:L07BCG_prog_compare}}
\end{figure*}

Combining the $178$ BCGs and the $291$ non-BCGs, the final local mixed sample is created. We run this process 10 times to get 10 sets of local mixed sample avoiding biases from selecting a single sample. In order to examine the effect of non-BCG properties, we compare the $469$ local mixed sample with the $178$ pure local BCGs within them. 

In Fig.~\ref{fig:L07BCG_prog_compare}, the dotted colour lines present the property distributions of our 10 sets of mixed samples at $z\sim 0$, and the magenta solid lines are for our one set of $178$ pure BCGs (the properties of 10 sets  pure BCGs are very similar, therefore we plot only one set pure BCGs to keep the plots clean). We find very little effect from the non-BCGs on the BCG properties, such that the mixed sample have very similar structures ($n$ and \re), \rff , stellar mass and SFR/sSFR as pure BCGs.  Note that the stellar mass distribution of the local mixed sample has a relatively evident offset from the pure BCGs by a factor of $0.08$ dex. Nevertheless, the uncertainty derived from the descendants of our selected non-BCG progenitors is no larger than $\sim 0.1$ dex for all the properties we explore.

\subsubsection{Can Contaminants erase BCG Evolution?}
\label{sec:BCGevo_intrinsic}

In Section~\ref{sec:nocontam_highz} we find that the effect of contamination from non-BCG progenitors is very small on BCG progenitor stellar mass and size ($< 0.1$ dex), but is more evident on SFR/sSFR by increasing them by a factor of $\sim 0.4$ dex. In Section~\ref{sec:nocontam_local} we demonstrate that the  effect of  the descendants of non-BCG progenitors is no more than $\sim 0.1$ dex on local BCG structure, stellar mass or SFR/sSFR properties. In this section, we will examine whether the uncertainty introduced by both non-BCG progenitors and local non-BCGs will erase the BCG evolution since $z\sim 2$ and whether the BCG evolution we find by our method is intrinsic.

The properties of our 38 selected progenitors are plotted in Fig.~\ref{fig:L07BCG_prog_compare} in blue solid lines with shadow. Note that there is one progenitor has very bad original CANDELS UDS image which results in unreliable fitting result (see Fig.~\ref{fig:fits_prog_CANDL}). Therefore, we do not take into account its shape, size, and morphology in our discussions. Comparing the properties of our high-$z$ progenitors with the properties of the $z\sim 0$ mixed sample (colour dotted lines), we find that BCG evolution is evident since $z\sim 2$ even if uncertainties are taken into account. We discuss this for stellar mass, SFR/sSFR and size specifically below. 

We find that even if the  BCG mass growth  decreases  when the $0.06$ dex uncertainty from non-BCG progenitors and the $0.08$ dex uncertainty from local non-BCGs are considered, the mass build-up in BCGs remains clear, growing by a factor of $0.24$ dex over $z=0 - 2$. The systematic contamination cannot erase the change of BCG SFR/sSFR either since the difference in SFR/sSFR  between high-$z$ progenitors and their local counterparts ($\sim 1.8$ dex for SFR; $\sim 2.2$ dex for sSFR) is much larger than the $0.4$ dex uncertainty from non-BCG progenitors. In respect of effective radius, $0.4$ dex size growth still remains even if  the $0.1$ dex systematic contamination from non-BCG progenitors is considered. 

In Section~\ref{sec:nocontam_highz} we find that our selected progenitors whose SFR/sSFR is less than the median value are more likely to be the true BCG progenitors. Since a low-SFR/-sSFR subsample may be more likely the true BCG progenitors,  we examine their properties specifically. In Fig.~\ref{fig:L07BCG_prog_compare}, the grey dashed lines with shadow represents the property distribution of our selected progenitors with low star formation rate whose $\log sSFR < -9.87$. These lower star forming systems have a much lower SFR and sSFR than the entire selected progenitor sample (by design), and are slightly more compact, more concentrated, and more massive. Nevertheless,  the  evolution in our selections from $z\sim 2$ to $z\sim 0$ remains statistically evident. 

In all, we demonstrate that BCG evolution based on our selection of high-$z$ progenitors and the local descendants must intrinsically be true. The uncertainties introduced by the contaminant non-BCG progenitors and local non-BCGs have relatively little effect, and cannot account for the evident evolution since $z\sim 2$. Even considering the low-sSFR subsample of high-$z$ selected progenitors, the evolution we find for BCGs remains. Since there is no good way to separate true BCG progenitors from our high-$z$ non-BCG progenitors in observations, the main results in the following sections are based on the entire $38$ selected progenitors and the 10 sets of local mixed samples. 


\section{BCG Evolution since $z\sim 2$}
\label{sec:results}

\begin{figure*}
\centering{ 
\subfigure{
\centering{\includegraphics[scale=0.55]{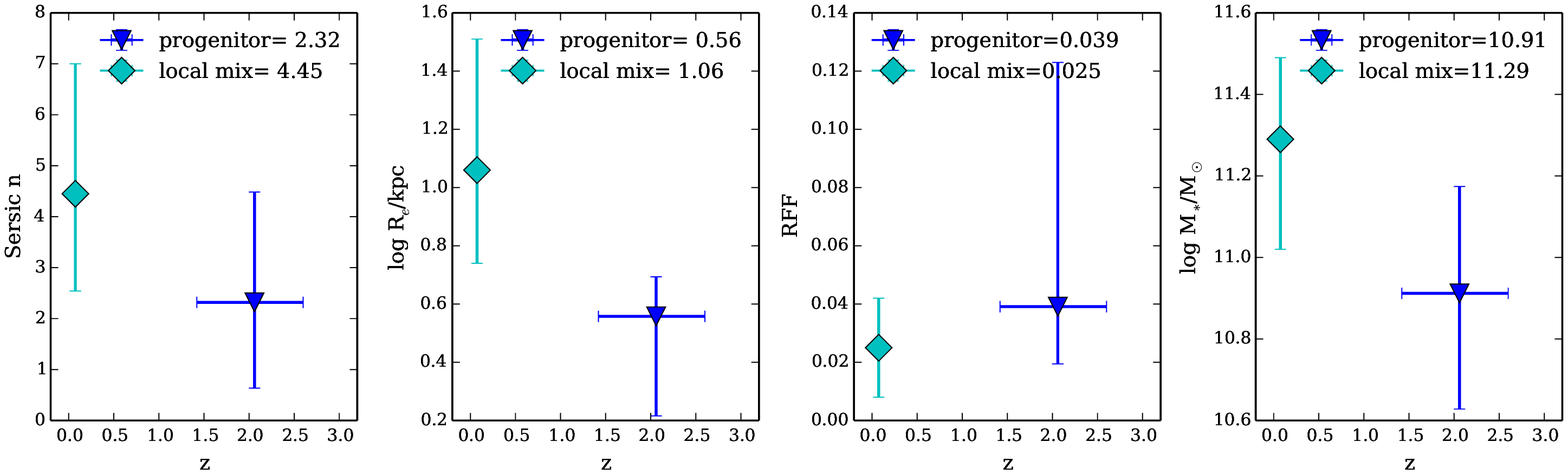}}}  
\subfigure{
\centering{\includegraphics[scale=0.55]{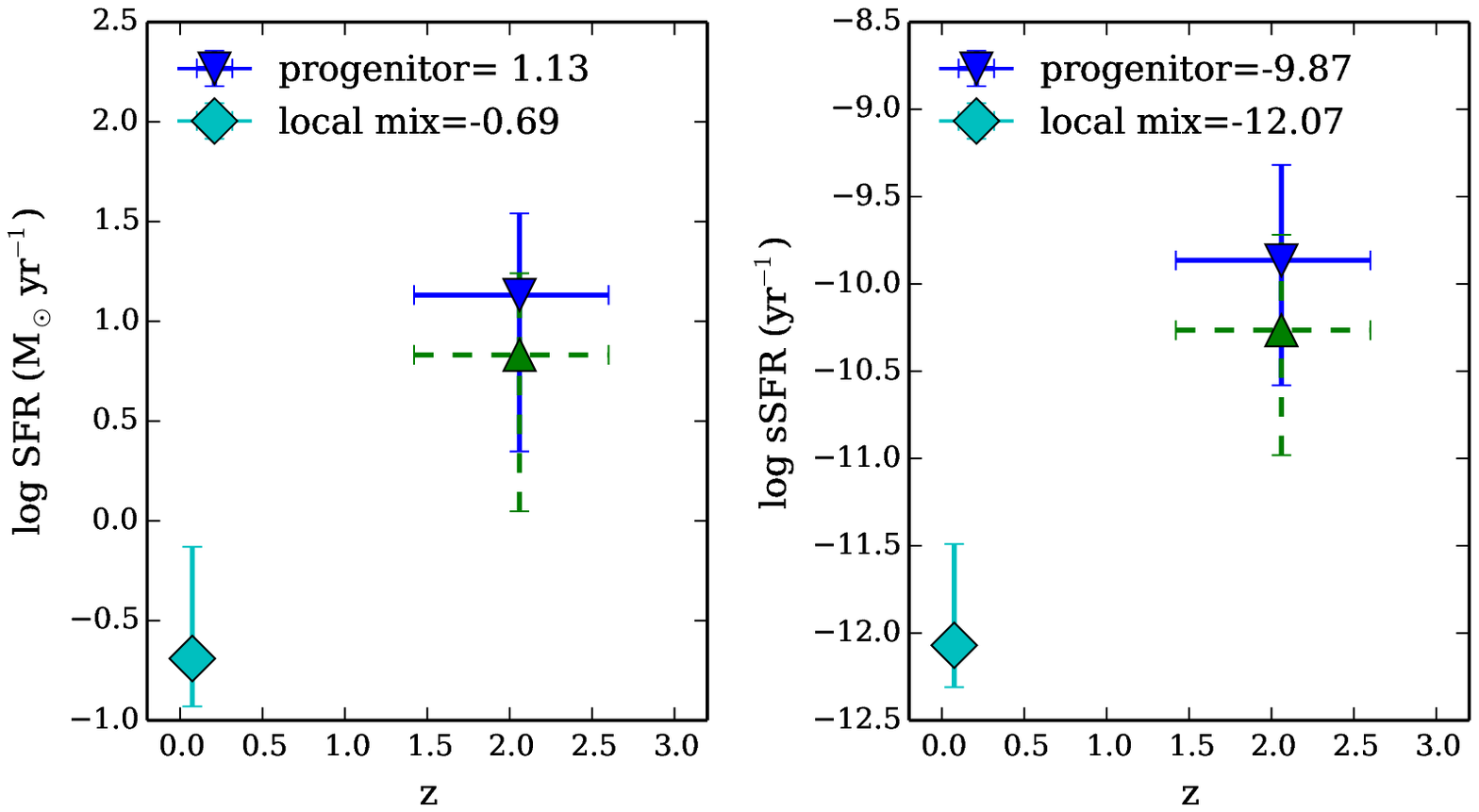}}}
\caption{The evolution of BCG properties as a function of redshift. The cyan diamonds show the median value of each property for the local descendants (obtained by averaging the medians of the 10 sets of local mixed samples simulated to $z=2$). The error bars are the 84 and 16 percentiles ($\sim 1\sigma$ of the distributions shown in Fig.~\ref{fig:L07BCG_prog_compare}), also averaged for the 10 sets local mixed samples. The blue triangle in each panel presents the median value of each property for our 38 high-$z$ progenitors. Note that the green triangles with dashed error bars in the panel of SFR and sSFR show the  median value with taking into account the difference between true BCG progenitors and our selected sample from the simulation. Clear evolution between $z\sim 2$ and $z\sim 0$ is observed for all the BCG properties presented (i.e., shape, size, morphology, stellar mass and SFR/sSFR). Note that the error bars represent the width of the distributions, and not the error in the median values, which are given in the text.}
\label{fig:L07BCG_prog_compare_avg}}
\end{figure*}

In order to explore BCG evolution since $z\sim 2$, we have selected 38 progenitors at $1\lesssim z\lesssim 3$ from the  CANDELS UDS and  created 10 sets of their local counterparts from SDSS DR7 as explained in detail in Section~\ref{sec:progselect}. We have demonstrated in  Section~\ref{sec:contam_effect} that the evolution between these two samples can  represent the BCG evolution from $z\sim 2$ to $z\sim 0$.  In this section using  Fig.~\ref{fig:L07BCG_prog_compare} and Fig.~\ref{fig:L07BCG_prog_compare_avg},  we  describe in detail the evolution of BCG structure (\sersic\ index  and effective radius), morphology (visual morphology and \rff), stellar mass and SFR/sSFR. Fig.~\ref{fig:L07BCG_prog_compare} presents the distributions of galaxy properties. Specifically, our main results of BCG evolution is shown by the blue solid lines with shadow (i.e., selected progenitors) and the colour dotted lines (i.e., local descendants). 

Fig.~\ref{fig:L07BCG_prog_compare_avg} explicitly illustrates how the BCG properties vary as a function of redshift. In Fig.~\ref{fig:L07BCG_prog_compare_avg}, the cyan diamond shows the mean value of each property, at our two different redshifts, by averaging the median value of the 10 sets constructed from the local samples (see Section~\ref{sec:progselect}). The error bars are the 84 and 16 percentiles ($\sim 1\sigma$) of each property distribution which are from averaging the error bars of 10 sets from the local samples. The median redshift of our local descendants is $0.074$. The blue triangle presents the median property value of our selected high-$z$ progenitors. The error bars are also the 84 and 16 percentiles of each property distribution. Our $38$ selected progenitors distribute around $z=2.06$. In the following, we call our selected high-$z$ systems the BCG progenitors, and call their local counterparts BCGs, for simplicity. 


\subsection{Structure Evolution}

Since the photometric images from the CANDELS survey are at high resolution, we can examine the   structures of the galaxies in our sample by fitting their light profiles. We use the pipeline of GALAPAGOS and GALFIT to fit each galaxy's profile with a single-\sersic\ model.  The sky value in the CANDELS imaging is determined by GALAPAGOS, and for the simulated BCG images the sky is fixed as the sky value obtained from the CANDELS sky patch used in the simulation. We analyse the behaviour of the two structural parameters derived from the best-fitting single-\sersic\ model. They are the \sersic\ index $n$, and the effective radius \re, which provides information on the intrinsic structural properties of these galaxies. 


\subsubsection{\sersic\ index $n$}
\label{sec:n_evolution}
The \sersic\ index $n$ measures the concentration of the light profile, with larger $n$ values corresponding to higher concentrations. The first panel in the upper row of Fig.~\ref{fig:L07BCG_prog_compare} clearly shows that the high-$z$ BCG progenitors have, statistically, much smaller values of $n$ than their local descendants. About $55$\% of BCG progenitors have $n$ smaller than $2.5$, which we define as late-type  galaxies. In contrast, less than $20$\% of the local BCGs have $n < 2.5$. A K-S test indicates that the difference in the \sersic\ index $n$ is significant at the $4.2\sigma$ level. The first panel in the upper row of Fig.~\ref{fig:L07BCG_prog_compare_avg} shows that the median $n$ of BCG progenitors is $2.32^{+0.44}_{-0.34}$, while the median $n$ of their local descendants is $4.45^{+0.15}_{-0.11}$. 

In previous work, \citet{Buitrago13} also find an enormous change for galaxy structures with cosmic time. They find that at $z\sim 2$, $\sim 70$\% of the massive galaxy population have  late-type \sersic\ profiles ($n<2.5$), while early-type galaxies ($n>2.5$) have been the predominant morphological class for massive galaxies since only $z\sim 1$. Our result suggests that the shape evolution is also true for the most massive galaxy population, the BCGs.

\subsubsection{Effective radius \re}
The effective radius \re\ is a measurement of the size of the light distribution. The second panel in the upper row of Fig.~\ref{fig:L07BCG_prog_compare} shows the distribution of $\log$\re\ for the high-$z$ BCG progenitors (blue solid lines with shadow), and their local descendants (dotted colour lines). It is clear that the BCG progenitors at $z \sim 2$ are much more compact than their descendants at $z\sim 0$. Almost all the high-$z$ progenitors ($> 90$\%) have radii smaller than \re$ \sim 6.3$ kpc, while there are $\sim 80$\% of local BCGs whose radii are larger than this value. The difference in $\log$\re\ distribution is very significant, based on a K-S test. The second panel in the upper row of Fig.~\ref{fig:L07BCG_prog_compare_avg} shows that the median radius of local BCGs is $11.5$ kpc (i.e., $\log$\re $=1.06^{+0.03}_{-0.02}$), which is a factor of $\sim 3.2$ larger than the size of the high-$z$ BCG progenitors ($\log$\re $=0.56^{+0.03}_{-0.07}$).   This is also similar to what is found when just selecting massive galaxies at high and low redshifts (\citealt{Buitrago13}).

\citet{Laporte13}  investigated the size growth of BCGs by using a suite of nine high-resolution dark matter-only simulations of galaxy clusters in a $\Lambda$CDM universe tracing a $z=2$ population of quiescent elliptical galaxies to $z=0$. They found that BCGs grow on average in size by a factor of $5-10$. This is much faster than the size growth we find  from observational data, such that BCGs grow in size  only by a factor of $3.2$ since $z\sim 2$. \citet{Laporte13} set the sizes of their high-$z$ galaxies according to the observed size-mass relation for $z \sim 2$ massive quiescent galaxies which have a steeper size--mass relation and experience faster size evolution (e.g., \citealt{Trujillo07}; \citealt{Buitrago08}; \citealt{vanderWel14}). In contrast, a large fraction of the BCG progenitors in our study are \sersic\ defined late-type galaxies (median $n=2.32$, see Section~\ref{sec:n_evolution}) which have slower size evolution (e.g., \citealt{Buitrago08}; \citealt{Bruce12}; \citealt{vanderWel14}). Therefore, it is not surprise that the size growth in \citet{Laporte13} is larger than our results. Since \citet{Buitrago13} demonstrate in observations that the late-type galaxies ($n<2.5$) dominate the massive galaxy population at $z>1$ (see also \citealt{Bruce12}), simulations need further improvement on exploring the size evolution of massive galaxies.

\subsection{Morphological Evolution}

The single-\sersic\ model is a generally reasonably good description of the local BCG light profiles since the majority of them are early-type galaxies. However, the high-$z$ BCG progenitors may be more complicated, with distorted features, or star forming regions and spiral arms, due to an intense early evolutionary phase. Therefore, inspection of the residuals that remain after subtracting the best-fit \sersic\ model is valuable for understanding whether a galaxy has a symmetric profile, or is in merger/star forming state.

We first carry out a visual inspection of the residual images which can generally give a good feel of whether the profiles of BCG progenitors and their local descendants are smooth or distorted. Then we demonstrate the quantitative differences by using the  objective diagnostic of \rff.

\subsubsection{Visual Inspect of Residual Images}
\label{sec:visual_morph}

\begin{figure*}
\centering{\includegraphics[scale=8.5]{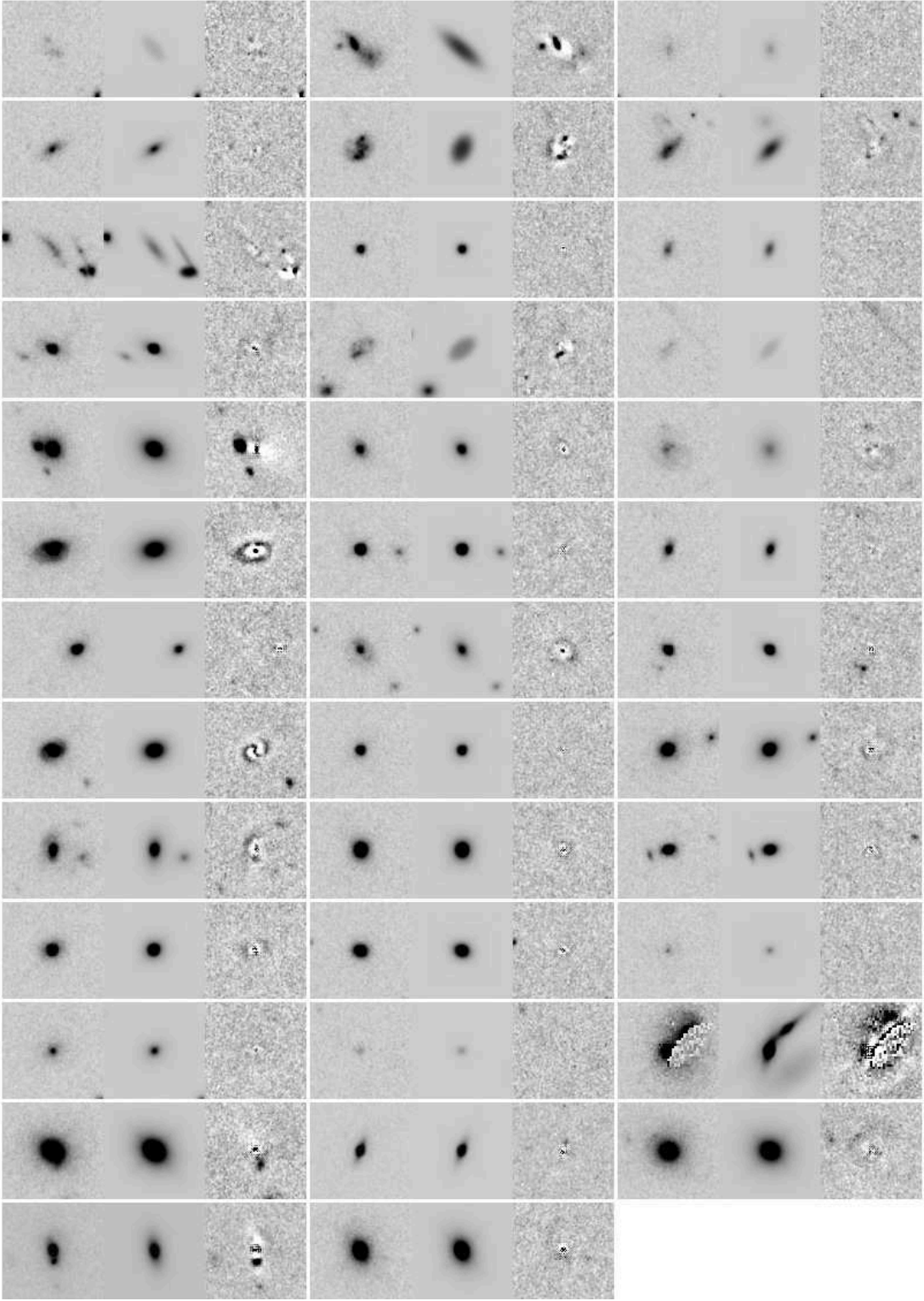}}
\caption{Single-\sersic\ fits of the $38$ BCG progenitors selected by our method from the CANDELS UDS. Each row presents three BCG progenitors, for each of which the left panel is the original image within the CANDELS UDS, the middle panel is the best-fit model, and the right panel is the residual image. The scale of each image is $5.2"\times 5.0"$. The first galaxy from right in the third row from bottom has very bad original image which results in unreliable fitting result. We do not take it into account in our discussions. Inspecting the residual images, $68$\% of the BCG progenitors have regular light profiles, the majority of which can be fitted by a single-\sersic\ model. In contrast, the remaining $32$\% of the progenitors are asymmetric, distorted, or have a close nearby companion. These imply that at $z \sim 2$  many BCG progenitors are undergoing or will undergo  interactions  and mergers.}
\label{fig:fits_prog_CANDL}
\end{figure*}

\begin{figure*}
\centering{\includegraphics[scale=6.2]{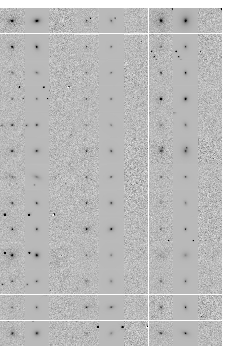}}   
\caption{Single-\sersic\ fits of $39$ local descendants which are simulated to $z=2$. They are randomly selected from one set of the local mixed BCG sample. As in Fig.~\ref{fig:fits_prog_CANDL}, each row presents three local descendants. The left panel is the original simulated image obtained from shifting the local SDSS BCG to $z=2$ by running the FERENGI code. The middle panel is the best-fit model, and the right panel is the residual image.  The scale of each image is $7.1"\times 7.3"$, corresponding to $59\times 61$ kpc$^2$ at $z=2$. It is clear that the local descendants have smooth and symmetric profiles, most of which can be well represented by a single-\sersic\ model.}
\label{fig:fits_localBCG}
\end{figure*}

Fig.~\ref{fig:fits_prog_CANDL} shows the single-\sersic\ fits for all the $38$ BCG progenitors selected by our method. Each row presents three BCG progenitors, each of which shows in the left panel the original image from the CANDELS UDS, the middle panel is the best-fitted single-\sersic\ model, and the right panel is the residual image. The first galaxy from right in the third row from bottom is excluded from our discussion since it has a bad original image.  

Inspecting the residual images, about $5$ of these systems have strong asymmetric/distorted profiles, or stretched structures, suggesting mergers are ongoing. Another $7$ progenitors show a close nearby object which implies that they might be undergo early stages of a merger.   The remaining  $25$ progenitors have regular profiles, $6$ of which show clear symmetric disc or spiral arms, while the others ($19$) can be well fit by a single-\sersic\ model. The single-\sersic\ fitting results in Fig.~\ref{fig:fits_prog_CANDL} indicate that more than half of the BCG progenitors seem to be in a quiescent evolutionary state which may already evolve  as elliptical galaxies. Nonetheless, there is still a large fraction of progenitors ($\sim 32$\%) which are undergoing, or will undergo, more intense interactions  at $z \lesssim 2$ with the responsible mechanism is most likely merging.

Fig.~\ref{fig:fits_localBCG} shows the single-\sersic\ fits of $39$ local descendants which are randomly selected from one set of our local mixed sample. As in Fig.~\ref{fig:fits_prog_CANDL}, each row shows three local descendants. The left panel is the original simulated image obtained from shifting the local BCG to $z=2$ by running the FERENGI code. The middle panel is the best-fitted single-\sersic\ model, and the right panel is the residual image. It is clear that all the local BCGs have smooth and symmetric profiles, most of which can be well represented by a single-\sersic\ model. None of these galaxies have an asymmetric or distorted morphology which can be found in the progenitor sample. This indicates that local BCGs are already well evolved into elliptical BCGs or cD galaxies (see also \citealt{Zhao15a}).  

\subsubsection{Residual Flux Fraction (\rff)}

Visual inspection of the residual images shows evident differences in light profile shapes between high-$z$ BCG progenitors and local BCGs, such that the high-$z$ ones are interacting  while the nearby ones already possess smooth profiles. In this section, we demonstrate this difference quantitatively through the \rff \ values whose calculation is in Section~\ref{sec:RFF_cal}. 

The \rff \ distributions, measured on the residual images of both the high-$z$ progenitors and the 10 sets of local descendants, are shown in the third panel in the upper row of Fig.~\ref{fig:L07BCG_prog_compare}. Local BCGs, whose residual images are visually clean with little obvious residuals, have a smaller \rff \  such that about $\sim 75$\% of them have $RFF \lesssim 0.03$. In contrast,  \rff \  of the BCG progenitors distributes towards larger values, indicating that a  fraction of them deviate further from the single-\sersic\ model. This is consistent with their light profiles. A K-S test shows a significant difference between these two \rff \  distributions at the level of $4.9 \sigma$. 

The difference of \rff \  between local BCGs and their progenitors is also shown in the third panel in the upper row of Fig.~\ref{fig:L07BCG_prog_compare_avg}, with the median \rff \  for BCGs at $z\sim 0$ being $0.025^{+0.001}_{-0.001}$, and for the high-$z$ progenitors it is $0.039^{+0.017}_{-0.004}$. Note that the \rff\ distribution of high-$z$ progenitors has a significant tail towards high values. From visual inspection on Fig.~\ref{fig:fits_prog_CANDL}, a fraction of progenitors are merging or have very close companions, creating a variety of unsmooth galaxy profiles. These profiles may result in \rff \  scattering towards larger values.

\subsection{Stellar Mass Evolution}
\label{sec:smass_SFR}

In this section, we probe BCG mass growth since $z \sim 2$. Since star formation is one potential mechanism for the increase of BCG stellar masses, we also compare SFR and sSFR of BCG progenitors at $z \sim 2$ and their descendants at $z \sim 0$ to determine how  star formation contributes to BCG mass growth.

\subsubsection{Stellar Mass Growth}
\label{sec:smass_growth}
The fourth panel in the upper row of Fig.~\ref{fig:L07BCG_prog_compare_avg} illustrates the average stellar mass difference of high-$z$ BCG progenitors and their $z\sim 0$ descendants. BCG stellar mass has grown by a factor of $\sim 2.5$ since $z \sim 2$ from $\log M_* = 10.91^{+0.05}_{-0.06}$ to $\log M_* = 11.29^{+0.01}_{-0.02}$. The fourth panel in the upper row of Fig.~\ref{fig:L07BCG_prog_compare} shows that  at $z \sim 2$ about 80\% of BCG progenitors have stellar masses smaller than $10^{11} M_\odot$, while in local universe $\sim 90$\% of descendants have grown into massive galaxies with masses larger than $10^{11} M_\odot$. The mass distributions of high-$z$ progenitors and local BCGs have significant differences at the $6.4 \sigma$ level as demonstrated by a K-S test.

Note that the stellar masses we use for the local sample are the MPA--JHU masses which are derived from the SDSS Petrosian magnitude. Studies, such as \citet{Bernardi13, Bernardi16} and \citet{DSouza15}, show that the SDSS Petrosian mass could underestimate the actual stellar mass especially for  massive galaxies. Based on the best-fit models of our single-\sersic\ fits which are possibly more appropriate to represent the light profiles of massive galaxies, we derive  the stellar masses for our local descendants by using their best-fit \sersic\ magnitudes. We find that  the Petrosian stellar mass is statistically smaller than the \sersic\ stellar mass by a factor of $0.3$ dex. Therefore, by using the \sersic\ stellar mass, the BCG  mass growth between $z\sim 2$ and $z\sim 0$ is larger by $0.3$ dex.

The evolution of BCG stellar mass  predicted in simulations can be examined by comparing the stellar masses of our selected $3780$ true BCG progenitors at $z=2$ with the mass of their BCG descendants at $z=0$. We find that the $z=0$ BCG stellar mass is about $5$ times larger than their $z=2$ progenitors. There is more BCG growth in simulations than in the observational results.  This offset between simulation and observation may be due to the higher galaxy stellar mass predicted in \citet{DeLB07} at low redshifts which is discussed in more detail in \citet{Lin13}. \citet{Laporte13} also predicts BCG evolution in simulations, but by adopting the dark matter-only simulations of galaxy clusters. They claim that the stellar mass of BCGs increase by a factor of $2-3$ since $z\sim 2$, which is consistent with our results.

\subsubsection{SFR and sSFR}
\label{sec:SFR_sSFR}

The left panel in the lower row of Fig.~\ref{fig:L07BCG_prog_compare} presents two clearly distinct distributions of SFR for BCG progenitors and local descendants. Almost all the BCG progenitors are forming more than $1$ $ M_\odot$ yr$^{-1}$ through star formation. The same panel in Fig.~\ref{fig:L07BCG_prog_compare_avg} indicates that their median SFR is  $13.5 ^{+4.3}_{-2.3}$ $M_\odot$ yr$^{-1}$. In the local universe, as BCGs have become  elliptical BCGs or cD galaxies, $\sim 85$\% of them have very low SFR that produce less than one solar mass per year. The median value of SFR for local BCGs from Fig.~\ref{fig:L07BCG_prog_compare_avg} is $ 0.20^{+0.03}_{-0.01}$ $M_\odot$ yr$^{-1}$.

The right panel in the lower row of Fig.~\ref{fig:L07BCG_prog_compare} shows the distributions of sSFR for BCG progenitors and their $z\sim 0$ descendants. Like SFRs, the sSFRs clearly separate  as well. From the right panel in the lower row of Fig.~\ref{fig:L07BCG_prog_compare_avg}, the high-$z$ BCG progenitors have a much higher sSFR concentrating on $\log \rm sSFR = -9.90^{+0.12}_{-0.14}$, while the sSFR of their descendants distributes around a very small value of  $\log \rm sSFR = -12.10 ^{+0.03}_{-0.01}$. 

\subsubsection{$M_*$--SFR relationship}

\begin{figure}
\centering{ 
\subfigure{
\centering{\includegraphics[scale=0.8]{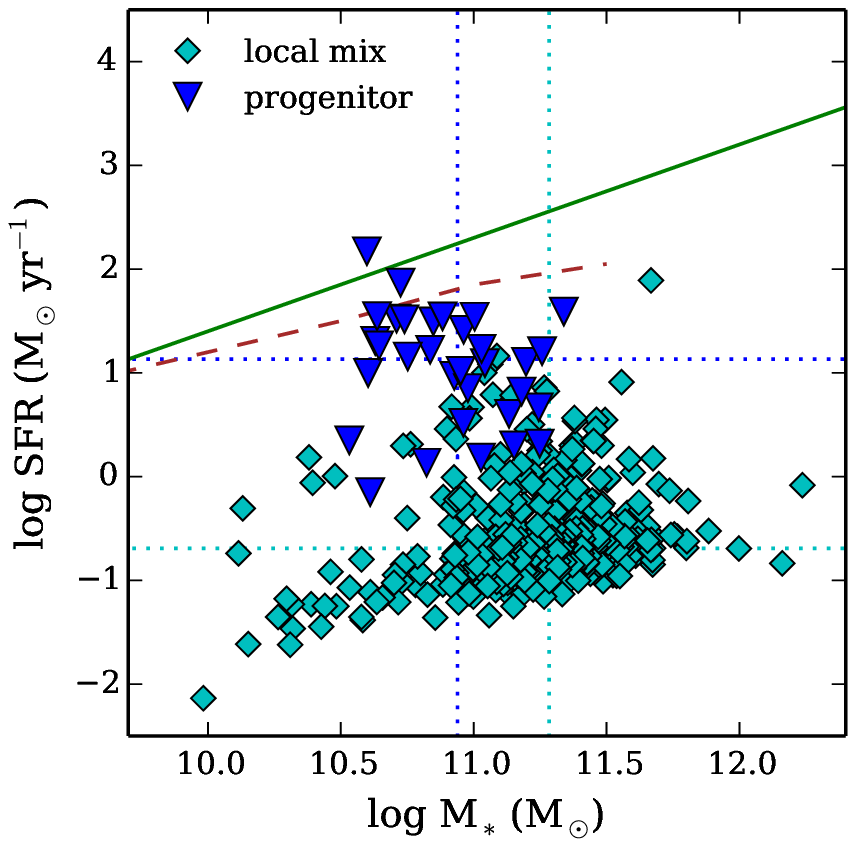}}}  
\subfigure{
\centering{\includegraphics[scale=0.76]{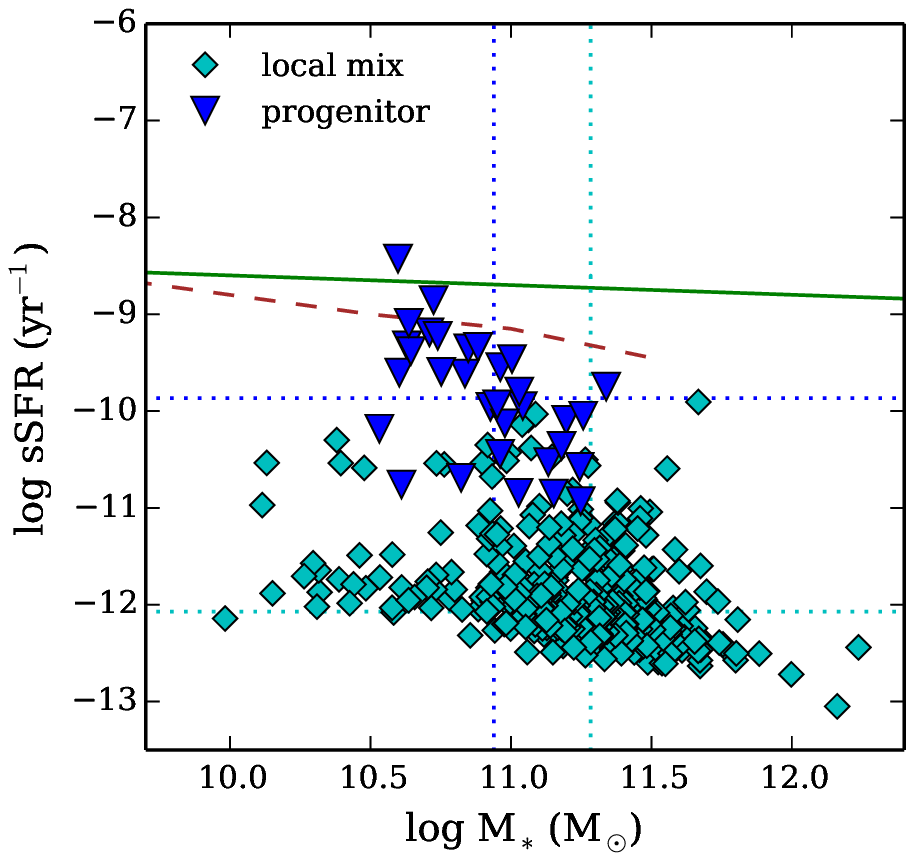}}}
\caption{\textit{Upper} panel: The $M_*$--SFR relation in log units for our 38 selected progenitors (blue triangles), and one set of local mixed sample (cyan diamonds).  The vertical dotted lines show the median values of log stellar mass, and the horizontal dotted lines indicate the median $\log SFR$. The green solid line is the relation found in \citet{Daddi07} for star-forming galaxies at $z=2$. The brown dashed line is the $M_*$--SFR relation from \citet{Bauer11} for star-forming galaxies at $2.0 < z < 2.5$.  \textit{Lower} panel: The $M_*$--sSFR relation in log units. The markers and lines have the  same meaning as in the \textit{upper} panel. It is clear that the BCG progenitors at $z\sim 2$ have a higher SFR and sSFR distributing separately from majority of their quiescent local descendants in either the $M_*$--SFR  or $M_*$--sSFR diagram, although the SFR and sSFR of BCG progenitors are lower than that of the general star-forming galaxies.}
\label{fig:smass_SFR}}
\end{figure}

Fig.~\ref{fig:smass_SFR} illustrates the $M_*$--SFR relation (\textit{upper} panel) as well as the $M_*$--sSFR relation (\textit{lower} panel) for our high-$z$ BCG progenitors and  their local descendants. Since the SFR/sSFR distributions are similar between the 10 sets of local descendants (see the lower row of Fig.~\ref{fig:L07BCG_prog_compare}), we plot only one set of the local sample in Fig.~\ref{fig:smass_SFR} to keep the figure clear. In each panel, the green solid line is the relation found in \citet{Daddi07} for star-forming galaxies at $z=2$, and the brown dashed line is the $M_*$--SFR relation from \citet{Bauer11} for star-forming galaxies at $2.0 < z < 2.5$.

It is evident that the BCG progenitors have lower SFR/sSFR values than the general star-forming galaxies, but still distribute in a relatively higher SFR/sSFR region differentiating from the majority of their local quiescent descendants. This implies that the BCG progenitors at $z \sim 2$ already passed through their most active star-forming phase, and have begun a quiescent phase. Nevertheless their less-intense star formation still keeps them in a relatively higher SFR/sSFR. In the local universe, however, their descendants have already long been quenched. Moreover, the morphologies of BCG progenitors (see Section~\ref{sec:visual_morph}) have no strong correlation with their SFRs or stellar masses.

\section{Discussion}
\label{sec:discussion}


\subsection{Mechanisms driving BCG mass growth}
\label{sec:discuss_mechanism}

The processes that increase the stellar masses and sizes of  massive galaxies are still an open question. There are two primary mechanisms: star formation and merging. Mergers are important since massive galaxies very likely form through the merging together of smaller galaxies in the hierarchical picture of galaxy formation. Star formation is also essential for massive galaxies in building up stellar mass, particularly at high redshifts where massive galaxies experience a much higher SFR than in the local universe (e.g., \citealt{vanDokkum04}; \citealt{Papovich06}; \citealt{Ownsworth12}). In this section, we will discuss the contribution of these two processes to the evolution of BCGs and their importance at different epochs.

Our study shows that BCG progenitors at $z\sim 2$ have a relatively high SFR, and a large fraction of them have either close companions or an asymmetric and distorted morphology.  These results suggest that  both star formation and (major) mergers may be key mechanisms in BCG evolution at $z \sim 2$. Here we carry out a simple estimate to determine how much these two mechanisms contribute to the BCG mass growth at high redshift. Note that the in-situ stellar mass of BCG progenitors at $z\sim 2$ already accounts for $\sim 40$\% of the total mass of BCGs at $z\sim 0$.

With the assumption that the SFR of our BCG progenitors is constant over $z=1-2$, and taking into account the $0.4$ dex uncertainty in SFR from contaminants, the mass increase during this period via star formation is $0.07 - 0.18M_{*,\rm z=0}$ where $M_{*,\rm z=0}$ is the stellar mass of BCGs at $z\sim 0$. On the other hand, we estimate the possible BCG mass growth through mergers by employing the major merger rate for massive galaxies at high redshifts. \citet{Conselice08} use the CAS parameters (structural concentration, asymmetry and clumpiness) to estimate major merger rates for galaxies at $1<z<3$. Since the median redshift of our BCG progenitors is $z\sim 2$ and 80\% of them have stellar mass less than $10^{11} M_\odot$, we use their major merger rate for galaxies with stellar masses $>10^{10}M_\odot$ at  $z=2$. Assuming the major merger rate is constant over $z=1-2$, we find that BCG mass growth is about $0.12 M_{*,\rm z=0}$ through major mergers. A similar mass increase is found by employing the major merger rate of \citet{Hopkins10}, such that for $z=2$ massive galaxies ($M_* > 10^{10}M_\odot$)   $0.09 M_{*,\rm z=0}$ is built up through merging with other objects whose mass ratios are $>1/3$. The star formation and major mergers thus seem to contribute equally to BCG mass build-up at high redshifts.


Our results show that the local BCGs are quite quiescent, where the mass added via star formation is only $0.2 M_\odot$ per year on average. Since the SFR of massive galaxies decreases quickly with cosmic time (e.g., \citealt{Daddi07}; \citealt{vanDokkum10}; \citealt{Ownsworth12}; \citealt{Ownsworth14}), the contribution from star formation to BCG mass growth since $z\sim 1$ should be very small. By using state-of-the-art, cosmological, semi-empirical models, \citet{BuchanShankar16} show that  star formation cannot explain the full evolution of massive galaxies, and massive galaxies could have indeed assembled most of their final masses via late mergers. Observationally, by studying the number of mergers onto BCGs, as well as the mass ratio of infalling companions, \citet{Burke13} find that both major and minor mergers are common at $z\sim 1$, and cause a significant BCG mass growth. At much lower redshifts, some observational studies conclude that minor mergers dominate mass growth, and the rarity of major mergers  (e.g., \citealt{Liu09}; \citealt{EP12}).  Others point out that some BCGs continue to grow through major mergers at $z\sim 0$. Nevertheless, merger (either major or minor) is the dominant process at $z\lesssim 1$.


\subsection{Links with BCG evolution at $z<1$}
In this work, we extend the observational study of BCG structural evolution and mass growth to $z \sim 2$.

In observations, BCG size evolution has been explored at $z<1$. By tracing host halo masses to link BCG progenitors and descendants, \citet{Shankar15} suggest a noticeable increase in BCG mean effective radius by a factor of $\gtrsim 2.5$ since $z\sim 1$. By comparing local WINGS BCGs with high-$z$ $HST$ BCGs whose host clusters span the same range of X-ray luminosity, \citet{Ascaso11} claim a BCG size growth of a factor of $\sim 2$ within the last 6 Gyr (since $z\sim 0.6$). These results indicate that about 60\% of the size growth of local BCGs has occurred at $z\lesssim1$. Considering the size increase in our study (by a factor of 3.2 from $z\sim 2$ ), it seems that BCG size increases only moderately during  $z=1-2$.  

Galaxy shape also reveals important information on galaxy evolution. We find that the \sersic\ index $n$ of BCGs has a clear evolution, such that BCG progenitors are consistent with \sersic\ late-type galaxies at $z\sim 2$, which evolve into local BCGs as early-type galaxies. Moreover, the morphology of our BCG progenitors  indicates that a fraction of them are undergoing morphological transformations at $z\sim 2$ through merging, or will undergo mergers at $z<2$. However, at $z<1$, \citet{Ascaso11} find that the shape of BCGs has not changed significantly after $z\sim 0.6$. Since the single \sersic\ model mainly represents the shape of the central bulge, it probably implies that the morphological transformation of BCG bulges is still going on at $z\sim 2$, and is complete before $z\sim 0.6$, during which mergers may play an important role. After that the  size and mass growth is focused on the outer regions of BCGs. More observational studies on the shape evolution of BCGs are needed during $z=0 - 1$ to determine if this scenario is likely.

Moreover, many studies explore the build-up of BCG stellar mass at $z\lesssim 1-1.5$ in observations. Some of them claim that there is little change in BCG mass since $z\sim 1$ (\citealt{BCM00}; \citealt{Whiley08}; \citealt{Collins09}). In contrast, other papers (e.g.,  \citealt{Lidman12}; \citealt{Lin13}; \citealt{Shankar15}; \citealt{Zhang15}) find a generally consistent BCG mass growth by a factor of $\sim 2$ over $z=0-1$. In Section~\ref{sec:discuss_mechanism}, we did a simple estimation of BCG mass growth from $z\sim 2$ to $z\sim 1$, reporting that, in this period, at most 18\% of the total mass of local BCGs will be added through star formation, and $\sim 12$\% via major mergers. Since SFR and major merger rate decrease with cosmic time (e.g., \citealt{Bridge10}; \citealt{Bluck12}; \citealt{Ownsworth14}), this mass growth is more likely an upper limit. Considering the stellar mass BCG progenitors already have at $z\sim 2$ ($\sim 40$\% of the total mass of local BCGs), our estimate shows that by $z\sim 1$ the BCG stellar mass will be no more than 70\% of the total mass at $z=0$, suggesting that there has to be an additional mass build-up in BCGs after $z\sim 1$. The BCG mass will increase by a factor of no less than  $\sim 1.4$ from $z\sim 1$ to $z\sim 0$.



Although we discuss the BCG evolution by combining our work over $z=0\sim 2$ with other studies at $z\lesssim 1$, it is dangerous to do so since the BCG progenitor selections we use are different. Homogeneous BCG data over large range of redshift from future observations is necessary for better understanding the BCG evolution since high redshifts.


\subsection{Comparison with massive galaxy evolution}

Many studies have examined the properties of massive galaxies at high redshifts, broadening our understanding of massive galaxy evolution over a large redshift range. Here we compare our results on BCGs with the evolution of massive galaxies over $z=0 - 2$. Since constant number density is applied in our study, the comparison is carried out  with papers which also use constant number density to trace massive galaxies at different redshifts.

\citet{vanDokkum10} study the growth of massive galaxies from $z=2$ using a fixed number density selection of $2 \times 10^{-4}$ Mpc$^{-3}$.  They find that at this number density the stellar mass of galaxies has increased by a factor of $\sim 2$, and size has grown by a factor of $\sim 4$ since $z=2$.  They verify that their results are not sensitive to the exact number density by repeating key parts of the analysis for a number density of  $1 \times 10^{-4}$ Mpc$^{-3}$. \citet{Ownsworth14} study the growth of massive galaxies from $z=3$ by adopting a fixed number density of $ \sim 10^{-4}$ Mpc$^{-3}$, similar to the one used in this paper. Their results show that the stellar mass of galaxies at $z\sim 0.3$ is $\sim 2.5$ times larger than their progenitors at $z\sim 2$, and the size of massive galaxies increases by a factor of $\sim 2.3$ by comparing the average galaxy size within the redshift bin $0.3<z<0.5$ with the bin at $2.0<z<2.5$.  Compared with  BCG stellar mass growth (a factor of $\sim 2.5$) and size growth (by a factor of $\sim 3.2$), the evolution of massive galaxies appears similar to the BCG evolution from $z\sim2$. 


Specifically, at high redshift,  we examine whether our selected BCG progenitors have different properties from normal massive galaxies which are in the same redshift and stellar mass range. The normal massive galaxies are selected from the CANDELS UDS catalogue whose redshifts and stellar masses have a similar distribution as our 38 selected BCG progenitors. We find that our BCG progenitors are very similar to the normal massive galaxies in many properties such as structure, morphology, and SFR/sSFR. This implies that the BCG progenitors do not show any specific differences with other massive galaxies at $z\sim 2$. Since local BCGs are different from the control samples of local non-BCGs which match in stellar mass, redshift and colour (see L07), BCG progenitors must experience some specific mechanism(s) at $z \lesssim 2$ (probably more minor mergers) which results in the specific properties of BCGs at $z\sim 0$. These mechanisms are likely responsible for the characteristic cD envelope observed in many local BCGs (\citealt{Zhao15a,Zhao15b}).

\section{Summary}
\label{sec:summary}

In this paper, we carry out a study of BCG evolution beyond $z=1$ to explore how structure, morphology and stellar mass of BCGs vary with cosmic time since $z\sim 2$. 

By proposing a BCG progenitor selection which identifies BCG progenitors as the most massive galaxies in the densest local environments, we select our BCG progenitor sample at $z\sim 2$ from the CANDELS UDS data. Testing our method in simulations we find that 45\% of our selected progenitors are true BCG progenitors. Although the high-$z$ progenitors selected by our method are a mixed sample of BCG and non-BCG progenitors, the properties of our high-$z$ progenitors can be used to trace BCG evolution because they are similar to the properties of the pure BCG progenitors within the sample. We use a constant number density of $10^{-4.06} h^3$Mpc$^{-3}$ to select our samples.  

At this density the descendants of the high-$z$ selected sample are taken from the SDSS DR7 galaxy catalogue.  To ensure the galaxy sample at $z\sim 0$ are the descendants of our selected progenitors, based on simulations, we construct a local  mixed sample which contains 38\% BCGs and 62\% non-BCGs. We demonstrate through several methods that the contamination from non-BCGs and non-BCG progenitors do not erase the intrinsic BCG evolution. Comparing properties between our high-$z$ BCG progenitors and their local descendants, we find a clear BCG evolution since $z\sim 2$ in structure, morphology and stellar mass.  Our major results on BCG evolution at $z \lesssim 3$ are:

\begin{itemize}
\item At $z\sim 2$, less than 50\% of the most massive galaxies in the densest environments are the true BCG progenitors.

\item Although the environmental density is not a strong tracer, the method we propose to identify BCG progenitors at $z\sim 2$ can be applied to observational data to derive BCG evolution since they have similar properties to the pure BCG progenitors.

\item The size of BCGs has grown by a factor of $\sim 3.2$ since $z\sim 2$. The BCG progenitor profiles are mainly \sersic\ late-type galaxies with median \sersic\ index of $n=2.3$, while their local BCG descendants are early-type galaxies whose median \sersic\ index is $n=4.5$.

\item The residual images after subtracting single \sersic\ fits illustrate that BCG progenitors at $z\sim 2$ are more distorted, whereas the local BCGs have smoother profiles. This difference in morphology is verified quantitatively by \rff \  measures, such that BCG progenitors have larger \rff \  values than their local counterparts. About $32$\% of BCG progenitors at $z\sim 2$ are  undergoing mergers, or will undergo mergers at $z<2$.

\item The stellar mass of BCGs has grown by a factor of $\sim 2.5$ since $z\sim 2$. The average SFR of BCG progenitors at $z\sim 2$ is still relatively high, at  $13.5$ $M_\odot$ yr$^{-1}$. In contrast, their local descendants are very quiescent, with an average SFR of only $ 0.2$ $M_\odot$ yr$^{-1}$. We find that over the $z=1 - 2$ period, star formation and merging contribute approximately equally to BCG mass growth. However, since the SFR decreases with time, merging must play a more important role in BCG assembly at $z\lesssim1$.

\item We find that BCG progenitors at high-$z$ are not significantly different than other galaxies of similar mass at the same redshift range. This suggests that the processes which differentiate BCGs from normal massive elliptical galaxies must occur at $z \lesssim 2$.

\end{itemize}

\section*{Acknowledgments}
We would like to thank Stuart Muldrew and Frazer Pearce for their helpful discussions. DZ's work is supported by a Research Excellence Scholarship from the University of Nottingham and the China Scholarship Council. A.M. acknowledges funding from the STFC and a European Research Council Consolidator Grant (P.I. R. McLure). AAS and CJC acknowledge financial support from the UK Science and Technology Facilities Council. This paper is partially based on SDSS data. Funding for SDSS-III has been provided by the Alfred P. Sloan Foundation, the Participating Institutions, the National Science Foundation, and the U.S. Department of Energy Office of Science. The SDSS-III web site is http://www.sdss3.org/. SDSS-III is managed by the Astrophysical Research Consortium for the Participating Institutions of the SDSS-III Collaboration including the University of Arizona, the Brazilian Participation Group, Brookhaven National Laboratory, Carnegie Mellon University, University of Florida, the French Participation Group, the German Participation Group, Harvard University, the Instituto de Astrofisica de Canarias, the Michigan State/Notre Dame/JINA Participation Group, Johns Hopkins University, Lawrence Berkeley National Laboratory, Max Planck Institute for Astrophysics, Max Planck Institute for Extraterrestrial Physics, New Mexico State University, New York University, Ohio State University, Pennsylvania State University, University of Portsmouth, Princeton University, the Spanish Participation Group, University of Tokyo, University of Utah, Vanderbilt University, University of Virginia, University of Washington, and Yale University.

\bibliographystyle{mnras}     
\bibliography{BCG_paper3}

\begin{thebibliography}{75}
\expandafter\ifx\csname natexlab\endcsname\relax\def\natexlab#1{#1}\fi

\bibitem[{{Ascaso} {et~al.}(2011){Ascaso}, {Aguerri}, {Varela}, {Cava},
  {Bettoni}, {Moles}, \& {D'Onofrio}}]{Ascaso11}
{Ascaso} B., {Aguerri} J.~A.~L., {Varela} J., {Cava} A., {Bettoni} D., {Moles}
  M., {D'Onofrio} M., 2011, \apj, 726, 69

\bibitem[{{Barden} {et~al.}(2012){Barden}, {H{\"a}u{\ss}ler}, {Peng},
  {McIntosh}, \& {Guo}}]{Barden12}
{Barden} M., {H{\"a}u{\ss}ler} B., {Peng} C.~Y., {McIntosh} D.~H., {Guo} Y.,
  2012, \mnras, 422, 449

\bibitem[{{Barden} {et~al.}(2008){Barden}, {Jahnke}, \&
  {H{\"a}u{\ss}ler}}]{Barden08}
{Barden} M., {Jahnke} K., {H{\"a}u{\ss}ler} B., 2008, \apjs, 175, 105

\bibitem[{{Bauer} {et~al.}(2011){Bauer}, {Conselice}, {P{\'e}rez-Gonz{\'a}lez},
  {Gr{\"u}tzbauch}, {Bluck}, {Buitrago}, \& {Mortlock}}]{Bauer11}
{Bauer} A.~E., {Conselice} C.~J., {P{\'e}rez-Gonz{\'a}lez} P.~G.,
  {Gr{\"u}tzbauch} R., {Bluck} A.~F.~L., {Buitrago} F., {Mortlock} A., 2011,
  \mnras, 417, 289

\bibitem[{{Bernardi} {et~al.}(2016){Bernardi}, {Meert}, {Sheth}, {Fischer},
  {Huertas-Company}, {Maraston}, {Shankar}, \& {Vikram}}]{Bernardi16}
{Bernardi} M., {Meert} A., {Sheth} R.~K., {Fischer} J.-L., {Huertas-Company}
  M., {Maraston} C., {Shankar} F., {Vikram} V., 2016, arXiv:1604.01036

\bibitem[{{Bernardi} {et~al.}(2013){Bernardi}, {Meert}, {Sheth}, {Vikram},
  {Huertas-Company}, {Mei}, \& {Shankar}}]{Bernardi13}
{Bernardi} M., {Meert} A., {Sheth} R.~K., {Vikram} V., {Huertas-Company} M.,
  {Mei} S., {Shankar} F., 2013, \mnras, 436, 697

\bibitem[{{Bertin} \& {Arnouts}(1996)}]{BA96}
{Bertin} E., {Arnouts} S., 1996, \aaps, 117, 393

\bibitem[{{Bluck} {et~al.}(2012){Bluck}, {Conselice}, {Buitrago},
  {Gr{\"u}tzbauch}, {Hoyos}, {Mortlock}, \& {Bauer}}]{Bluck12}
{Bluck} A.~F.~L., {Conselice} C.~J., {Buitrago} F., {Gr{\"u}tzbauch} R.,
  {Hoyos} C., {Mortlock} A., {Bauer} A.~E., 2012, \apj, 747, 34

\bibitem[{{Bridge} {et~al.}(2010){Bridge}, {Carlberg}, \&
  {Sullivan}}]{Bridge10}
{Bridge} C.~R., {Carlberg} R.~G., {Sullivan} M., 2010, \apj, 709, 1067

\bibitem[{{Brinchmann} {et~al.}(2004){Brinchmann}, {Charlot}, {White},
  {Tremonti}, {Kauffmann}, {Heckman}, \& {Brinkmann}}]{Brinchmann04}
{Brinchmann} J., {Charlot} S., {White} S.~D.~M., {Tremonti} C., {Kauffmann} G.,
  {Heckman} T., {Brinkmann} J., 2004, \mnras, 351, 1151

\bibitem[{{Bruce} {et~al.}(2012){Bruce}, {Dunlop}, {Cirasuolo}, {McLure},
  {Targett}, {Bell}, {Croton}, {Dekel}, {Faber}, {Ferguson}, {Grogin},
  {Kocevski}, {Koekemoer}, {Koo}, {Lai}, {Lotz}, {McGrath}, {Newman}, \& {van
  der Wel}}]{Bruce12}
{Bruce} V.~A., {Dunlop} J.~S., {Cirasuolo} M., {McLure} R.~J., {Targett} T.~A.,
  {Bell} E.~F., {Croton} D.~J., {Dekel} A., {Faber} S.~M., {Ferguson} H.~C.,
  {Grogin} N.~A., {Kocevski} D.~D., {Koekemoer} A.~M., {Koo} D.~C., {Lai} K.,
  {Lotz} J.~M., {McGrath} E.~J., {Newman} J.~A., {van der Wel} A., 2012,
  \mnras, 427, 1666

\bibitem[{{Bruzual} \& {Charlot}(2003)}]{BC03}
{Bruzual} G., {Charlot} S., 2003, \mnras, 344, 1000

\bibitem[{{Buchan} \& {Shankar}(2016)}]{BuchanShankar16}
{Buchan} S., {Shankar} F., 2016, \mnras, 462, 2001

\bibitem[{{Buitrago} {et~al.}(2008){Buitrago}, {Trujillo}, {Conselice},
  {Bouwens}, {Dickinson}, \& {Yan}}]{Buitrago08}
{Buitrago} F., {Trujillo} I., {Conselice} C.~J., {Bouwens} R.~J., {Dickinson}
  M., {Yan} H., 2008, \apjl, 687, L61

\bibitem[{{Buitrago} {et~al.}(2013){Buitrago}, {Trujillo}, {Conselice}, \&
  {H{\"a}u{\ss}ler}}]{Buitrago13}
{Buitrago} F., {Trujillo} I., {Conselice} C.~J., {H{\"a}u{\ss}ler} B., 2013,
  \mnras, 428, 1460

\bibitem[{{Burke} \& {Collins}(2013)}]{Burke13}
{Burke} C., {Collins} C.~A., 2013, \mnras, 434, 2856

\bibitem[{{Burke} {et~al.}(2015){Burke}, {Hilton}, \& {Collins}}]{Burke15}
{Burke} C., {Hilton} M., {Collins} C., 2015, \mnras, 449, 2353

\bibitem[{{Burke} {et~al.}(2000){Burke}, {Collins}, \& {Mann}}]{BCM00}
{Burke} D.~J., {Collins} C.~A., {Mann} R.~G., 2000, \apjl, 532, L105

\bibitem[{{Caon} {et~al.}(1993){Caon}, {Capaccioli}, \& {D'Onofrio}}]{Caon93}
{Caon} N., {Capaccioli} M., {D'Onofrio} M., 1993, \mnras, 265, 1013

\bibitem[{{Cid Fernandes} {et~al.}(2005){Cid Fernandes}, {Mateus}, {Sodr{\'e}},
  {Stasi{\'n}ska}, \& {Gomes}}]{Fernandes05}
{Cid Fernandes} R., {Mateus} A., {Sodr{\'e}} L., {Stasi{\'n}ska} G., {Gomes}
  J.~M., 2005, \mnras, 358, 363

\bibitem[{{Collins} {et~al.}(2009){Collins}, {Stott}, {Hilton}, {Kay},
  {Stanford}, {Davidson}, {Hosmer}, {Hoyle}, {Liddle}, {Lloyd-Davies}, {Mann},
  {Mehrtens}, {Miller}, {Nichol}, {Romer}, {Sahl{\'e}n}, {Viana}, \&
  {West}}]{Collins09}
{Collins} C.~A., {Stott} J.~P., {Hilton} M., {Kay} S.~T., {Stanford} S.~A.,
  {Davidson} M., {Hosmer} M., {Hoyle} B., {Liddle} A., {Lloyd-Davies} E.,
  {Mann} R.~G., {Mehrtens} N., {Miller} C.~J., {Nichol} R.~C., {Romer} A.~K.,
  {Sahl{\'e}n} M., {Viana} P.~T.~P., {West} M.~J., 2009, \nat, 458, 603

\bibitem[{{Conroy} {et~al.}(2007){Conroy}, {Wechsler}, \&
  {Kravtsov}}]{Conroy07}
{Conroy} C., {Wechsler} R.~H., {Kravtsov} A.~V., 2007, \apj, 668, 826

\bibitem[{{Conselice} {et~al.}(2011){Conselice}, {Bluck}, {Ravindranath},
  {Mortlock}, {Koekemoer}, {Buitrago}, {Gr{\"u}tzbauch}, \&
  {Penny}}]{Conselice11}
{Conselice} C.~J., {Bluck} A.~F.~L., {Ravindranath} S., {Mortlock} A.,
  {Koekemoer} A.~M., {Buitrago} F., {Gr{\"u}tzbauch} R., {Penny} S.~J., 2011,
  \mnras, 417, 2770

\bibitem[{{Conselice} {et~al.}(2008){Conselice}, {Rajgor}, \&
  {Myers}}]{Conselice08}
{Conselice} C.~J., {Rajgor} S., {Myers} R., 2008, \mnras, 386, 909

\bibitem[{{Daddi} {et~al.}(2007){Daddi}, {Dickinson}, {Morrison}, {Chary},
  {Cimatti}, {Elbaz}, {Frayer}, {Renzini}, {Pope}, {Alexander}, {Bauer},
  {Giavalisco}, {Huynh}, {Kurk}, \& {Mignoli}}]{Daddi07}
{Daddi} E., {Dickinson} M., {Morrison} G., {Chary} R., {Cimatti} A., {Elbaz}
  D., {Frayer} D., {Renzini} A., {Pope} A., {Alexander} D.~M., {Bauer} F.~E.,
  {Giavalisco} M., {Huynh} M., {Kurk} J., {Mignoli} M., 2007, \apj, 670, 156

\bibitem[{{De Lucia} \& {Blaizot}(2007)}]{DeLB07}
{De Lucia} G., {Blaizot} J., 2007, \mnras, 375, 2

\bibitem[{{D'Souza} {et~al.}(2015){D'Souza}, {Vegetti}, \&
  {Kauffmann}}]{DSouza15}
{D'Souza} R., {Vegetti} S., {Kauffmann} G., 2015, \mnras, 454, 4027

\bibitem[{{Dubinski}(1998)}]{Dubinski98}
{Dubinski} J., 1998, \apj, 502, 141

\bibitem[{{Edwards} \& {Patton}(2012)}]{EP12}
{Edwards} L.~O.~V., {Patton} D.~R., 2012, \mnras, 425, 287

\bibitem[{{Fakhouri} {et~al.}(2010){Fakhouri}, {Ma}, \&
  {Boylan-Kolchin}}]{FakhouriMaBoylan10}
{Fakhouri} O., {Ma} C.-P., {Boylan-Kolchin} M., 2010, \mnras, 406, 2267

\bibitem[{{Fontana} {et~al.}(2014){Fontana}, {Dunlop}, {Paris}, {Targett},
  {Boutsia}, {Castellano}, {Galametz}, {Grazian}, {McLure}, {Merlin},
  {Pentericci}, {Wuyts}, {Almaini}, {Caputi}, {Chary}, {Cirasuolo},
  {Conselice}, {Cooray}, {Daddi}, {Dickinson}, {Faber}, {Fazio}, {Ferguson},
  {Giallongo}, {Giavalisco}, {Grogin}, {Hathi}, {Koekemoer}, {Koo}, {Lucas},
  {Nonino}, {Rix}, {Renzini}, {Rosario}, {Santini}, {Scarlata}, {Sommariva},
  {Stark}, {van der Wel}, {Vanzella}, {Wild}, {Yan}, \& {Zibetti}}]{Fontana14}
{Fontana} A., {Dunlop} J.~S., {Paris} D., {Targett} T.~A., {Boutsia} K.,
  {Castellano} M., {Galametz} A., {Grazian} A., {McLure} R., {Merlin} E.,
  {Pentericci} L., {Wuyts} S., {Almaini} O., {Caputi} K., {Chary} R.-R.,
  {Cirasuolo} M., {Conselice} C.~J., {Cooray} A., {Daddi} E., {Dickinson} M.,
  {Faber} S.~M., {Fazio} G., {Ferguson} H.~C., {Giallongo} E., {Giavalisco} M.,
  {Grogin} N.~A., {Hathi} N., {Koekemoer} A.~M., {Koo} D.~C., {Lucas} R.~A.,
  {Nonino} M., {Rix} H.~W., {Renzini} A., {Rosario} D., {Santini} P.,
  {Scarlata} C., {Sommariva} V., {Stark} D.~P., {van der Wel} A., {Vanzella}
  E., {Wild} V., {Yan} H., {Zibetti} S., 2014, \aap, 570, A11

\bibitem[{{Furusawa} {et~al.}(2008){Furusawa}, {Kosugi}, {Akiyama}, {Takata},
  {Sekiguchi}, {Tanaka}, {Iwata}, {Kajisawa}, {Yasuda}, {Doi}, {Ouchi},
  {Simpson}, {Shimasaku}, {Yamada}, {Furusawa}, {Morokuma}, {Ishida}, {Aoki},
  {Fuse}, {Imanishi}, {Iye}, {Karoji}, {Kobayashi}, {Kodama}, {Komiyama},
  {Maeda}, {Miyazaki}, {Mizumoto}, {Nakata}, {Noumaru}, {Ogasawara}, {Okamura},
  {Saito}, {Sasaki}, {Ueda}, \& {Yoshida}}]{Furusawa08}
{Furusawa} H., {Kosugi} G., {Akiyama} M., {Takata} T., {Sekiguchi} K., {Tanaka}
  I., {Iwata} I., {Kajisawa} M., {Yasuda} N., {Doi} M., {Ouchi} M., {Simpson}
  C., {Shimasaku} K., {Yamada} T., {Furusawa} J., {Morokuma} T., {Ishida}
  C.~M., {Aoki} K., {Fuse} T., {Imanishi} M., {Iye} M., {Karoji} H.,
  {Kobayashi} N., {Kodama} T., {Komiyama} Y., {Maeda} Y., {Miyazaki} S.,
  {Mizumoto} Y., {Nakata} F., {Noumaru} J., {Ogasawara} R., {Okamura} S.,
  {Saito} T., {Sasaki} T., {Ueda} Y., {Yoshida} M., 2008, \apjs, 176, 1

\bibitem[{{Grogin} {et~al.}(2011){Grogin}, {Kocevski}, {Faber}, {Ferguson},
  {Koekemoer}, {Riess}, {Acquaviva}, {Alexander}, {Almaini}, {Ashby}, {Barden},
  {Bell}, {Bournaud}, {Brown}, {Caputi}, {Casertano}, {Cassata}, {Castellano},
  {Challis}, {Chary}, {Cheung}, {Cirasuolo}, {Conselice}, {Roshan Cooray},
  {Croton}, {Daddi}, {Dahlen}, {Dav{\'e}}, {de Mello}, {Dekel}, {Dickinson},
  {Dolch}, {Donley}, {Dunlop}, {Dutton}, {Elbaz}, {Fazio}, {Filippenko},
  {Finkelstein}, {Fontana}, {Gardner}, {Garnavich}, {Gawiser}, {Giavalisco},
  {Grazian}, {Guo}, {Hathi}, {H{\"a}ussler}, {Hopkins}, {Huang}, {Huang},
  {Jha}, {Kartaltepe}, {Kirshner}, {Koo}, {Lai}, {Lee}, {Li}, {Lotz}, {Lucas},
  {Madau}, {McCarthy}, {McGrath}, {McIntosh}, {McLure}, {Mobasher},
  {Moustakas}, {Mozena}, {Nandra}, {Newman}, {Niemi}, {Noeske}, {Papovich},
  {Pentericci}, {Pope}, {Primack}, {Rajan}, {Ravindranath}, {Reddy}, {Renzini},
  {Rix}, {Robaina}, {Rodney}, {Rosario}, {Rosati}, {Salimbeni}, {Scarlata},
  {Siana}, {Simard}, {Smidt}, {Somerville}, {Spinrad}, {Straughn}, {Strolger},
  {Telford}, {Teplitz}, {Trump}, {van der Wel}, {Villforth}, {Wechsler},
  {Weiner}, {Wiklind}, {Wild}, {Wilson}, {Wuyts}, {Yan}, \& {Yun}}]{Grogin11}
{Grogin} N.~A., {Kocevski} D.~D., {Faber} S.~M., {Ferguson} H.~C., {Koekemoer}
  A.~M., {Riess} A.~G., {Acquaviva} V., {Alexander} D.~M., {Almaini} O.,
  {Ashby} M.~L.~N., {Barden} M., {Bell} E.~F., {Bournaud} F., {Brown} T.~M.,
  {Caputi} K.~I., {Casertano} S., {Cassata} P., {Castellano} M., {Challis} P.,
  {Chary} R.-R., {Cheung} E., {Cirasuolo} M., {Conselice} C.~J., {Roshan
  Cooray} A., {Croton} D.~J., {Daddi} E., {Dahlen} T., {Dav{\'e}} R., {de
  Mello} D.~F., {Dekel} A., {Dickinson} M., {Dolch} T., {Donley} J.~L.,
  {Dunlop} J.~S., {Dutton} A.~A., {Elbaz} D., {Fazio} G.~G., {Filippenko}
  A.~V., {Finkelstein} S.~L., {Fontana} A., {Gardner} J.~P., {Garnavich} P.~M.,
  {Gawiser} E., {Giavalisco} M., {Grazian} A., {Guo} Y., {Hathi} N.~P.,
  {H{\"a}ussler} B., {Hopkins} P.~F., {Huang} J.-S., {Huang} K.-H., {Jha}
  S.~W., {Kartaltepe} J.~S., {Kirshner} R.~P., {Koo} D.~C., {Lai} K., {Lee}
  K.-S., {Li} W., {Lotz} J.~M., {Lucas} R.~A., {Madau} P., {McCarthy} P.~J.,
  {McGrath} E.~J., {McIntosh} D.~H., {McLure} R.~J., {Mobasher} B., {Moustakas}
  L.~A., {Mozena} M., {Nandra} K., {Newman} J.~A., {Niemi} S.-M., {Noeske}
  K.~G., {Papovich} C.~J., {Pentericci} L., {Pope} A., {Primack} J.~R., {Rajan}
  A., {Ravindranath} S., {Reddy} N.~A., {Renzini} A., {Rix} H.-W., {Robaina}
  A.~R., {Rodney} S.~A., {Rosario} D.~J., {Rosati} P., {Salimbeni} S.,
  {Scarlata} C., {Siana} B., {Simard} L., {Smidt} J., {Somerville} R.~S.,
  {Spinrad} H., {Straughn} A.~N., {Strolger} L.-G., {Telford} O., {Teplitz}
  H.~I., {Trump} J.~R., {van der Wel} A., {Villforth} C., {Wechsler} R.~H.,
  {Weiner} B.~J., {Wiklind} T., {Wild} V., {Wilson} G., {Wuyts} S., {Yan}
  H.-J., {Yun} M.~S., 2011, \apjs, 197, 35

\bibitem[{{Hartley} {et~al.}(2013){Hartley}, {Almaini}, {Mortlock},
  {Conselice}, {Gr{\"u}tzbauch}, {Simpson}, {Bradshaw}, {Chuter}, {Foucaud},
  {Cirasuolo}, {Dunlop}, {McLure}, \& {Pearce}}]{Hartley13}
{Hartley} W.~G., {Almaini} O., {Mortlock} A., {Conselice} C.~J.,
  {Gr{\"u}tzbauch} R., {Simpson} C., {Bradshaw} E.~J., {Chuter} R.~W.,
  {Foucaud} S., {Cirasuolo} M., {Dunlop} J.~S., {McLure} R.~J., {Pearce} H.~J.,
  2013, \mnras, 431, 3045

\bibitem[{{H{\"a}ussler} {et~al.}(2007){H{\"a}ussler}, {McIntosh}, {Barden},
  {Bell}, {Rix}, {Borch}, {Beckwith}, {Caldwell}, {Heymans}, {Jahnke}, {Jogee},
  {Koposov}, {Meisenheimer}, {S{\'a}nchez}, {Somerville}, {Wisotzki}, \&
  {Wolf}}]{Haussler07}
{H{\"a}ussler} B., {McIntosh} D.~H., {Barden} M., {Bell} E.~F., {Rix} H.-W.,
  {Borch} A., {Beckwith} S.~V.~W., {Caldwell} J.~A.~R., {Heymans} C., {Jahnke}
  K., {Jogee} S., {Koposov} S.~E., {Meisenheimer} K., {S{\'a}nchez} S.~F.,
  {Somerville} R.~S., {Wisotzki} L., {Wolf} C., 2007, \apjs, 172, 615

\bibitem[{{Hopkins} {et~al.}(2010){Hopkins}, {Bundy}, {Hernquist}, {Wuyts}, \&
  {Cox}}]{Hopkins10}
{Hopkins} P.~F., {Bundy} K., {Hernquist} L., {Wuyts} S., {Cox} T.~J., 2010,
  \mnras, 401, 1099

\bibitem[{{Hoyos} {et~al.}(2011){Hoyos}, {den Brok}, {Verdoes Kleijn},
  {Carter}, {Balcells}, {Guzm{\'a}n}, {Peletier}, {Ferguson}, {Goudfrooij},
  {Graham}, {Hammer}, {Karick}, {Lucey}, {Matkovi{\'c}}, {Merritt}, {Mouhcine},
  \& {Valentijn}}]{Hoyos11}
{Hoyos} C., {den Brok} M., {Verdoes Kleijn} G., {Carter} D., {Balcells} M.,
  {Guzm{\'a}n} R., {Peletier} R., {Ferguson} H.~C., {Goudfrooij} P., {Graham}
  A.~W., {Hammer} D., {Karick} A.~M., {Lucey} J.~R., {Matkovi{\'c}} A.,
  {Merritt} D., {Mouhcine} M., {Valentijn} E., 2011, \mnras, 411, 2439

\bibitem[{{Ilbert} {et~al.}(2005){Ilbert}, {Tresse}, {Zucca}, {Bardelli},
  {Arnouts}, {Zamorani}, {Pozzetti}, {Bottini}, {Garilli}, {Le Brun}, {Le
  F{\`e}vre}, {Maccagni}, {Picat}, {Scaramella}, {Scodeggio}, {Vettolani},
  {Zanichelli}, {Adami}, {Arnaboldi}, {Bolzonella}, {Cappi}, {Charlot},
  {Contini}, {Foucaud}, {Franzetti}, {Gavignaud}, {Guzzo}, {Iovino},
  {McCracken}, {Marano}, {Marinoni}, {Mathez}, {Mazure}, {Meneux}, {Merighi},
  {Paltani}, {Pello}, {Pollo}, {Radovich}, {Bondi}, {Bongiorno}, {Busarello},
  {Ciliegi}, {Lamareille}, {Mellier}, {Merluzzi}, {Ripepi}, \&
  {Rizzo}}]{Ilbert05}
{Ilbert} O., {Tresse} L., {Zucca} E., {Bardelli} S., {Arnouts} S., {Zamorani}
  G., {Pozzetti} L., {Bottini} D., {Garilli} B., {Le Brun} V., {Le F{\`e}vre}
  O., {Maccagni} D., {Picat} J.-P., {Scaramella} R., {Scodeggio} M.,
  {Vettolani} G., {Zanichelli} A., {Adami} C., {Arnaboldi} M., {Bolzonella} M.,
  {Cappi} A., {Charlot} S., {Contini} T., {Foucaud} S., {Franzetti} P.,
  {Gavignaud} I., {Guzzo} L., {Iovino} A., {McCracken} H.~J., {Marano} B.,
  {Marinoni} C., {Mathez} G., {Mazure} A., {Meneux} B., {Merighi} R., {Paltani}
  S., {Pello} R., {Pollo} A., {Radovich} M., {Bondi} M., {Bongiorno} A.,
  {Busarello} G., {Ciliegi} P., {Lamareille} F., {Mellier} Y., {Merluzzi} P.,
  {Ripepi} V., {Rizzo} D., 2005, \aap, 439, 863

\bibitem[{{Kauffmann} {et~al.}(2003){Kauffmann}, {Heckman}, {White}, {Charlot},
  {Tremonti}, {Brinchmann}, {Bruzual}, {Peng}, {Seibert}, {Bernardi},
  {Blanton}, {Brinkmann}, {Castander}, {Cs{\'a}bai}, {Fukugita}, {Ivezic},
  {Munn}, {Nichol}, {Padmanabhan}, {Thakar}, {Weinberg}, \&
  {York}}]{Kauffmann03}
{Kauffmann} G., {Heckman} T.~M., {White} S.~D.~M., {Charlot} S., {Tremonti} C.,
  {Brinchmann} J., {Bruzual} G., {Peng} E.~W., {Seibert} M., {Bernardi} M.,
  {Blanton} M., {Brinkmann} J., {Castander} F., {Cs{\'a}bai} I., {Fukugita} M.,
  {Ivezic} Z., {Munn} J.~A., {Nichol} R.~C., {Padmanabhan} N., {Thakar} A.~R.,
  {Weinberg} D.~H., {York} D., 2003, \mnras, 341, 33

\bibitem[{{Koekemoer} {et~al.}(2011){Koekemoer}, {Faber}, {Ferguson}, {Grogin},
  {Kocevski}, {Koo}, {Lai}, {Lotz}, {Lucas}, {McGrath}, {Ogaz}, {Rajan},
  {Riess}, {Rodney}, {Strolger}, {Casertano}, {Castellano}, {Dahlen},
  {Dickinson}, {Dolch}, {Fontana}, {Giavalisco}, {Grazian}, {Guo}, {Hathi},
  {Huang}, {van der Wel}, {Yan}, {Acquaviva}, {Alexander}, {Almaini}, {Ashby},
  {Barden}, {Bell}, {Bournaud}, {Brown}, {Caputi}, {Cassata}, {Challis},
  {Chary}, {Cheung}, {Cirasuolo}, {Conselice}, {Roshan Cooray}, {Croton},
  {Daddi}, {Dav{\'e}}, {de Mello}, {de Ravel}, {Dekel}, {Donley}, {Dunlop},
  {Dutton}, {Elbaz}, {Fazio}, {Filippenko}, {Finkelstein}, {Frazer}, {Gardner},
  {Garnavich}, {Gawiser}, {Gruetzbauch}, {Hartley}, {H{\"a}ussler},
  {Herrington}, {Hopkins}, {Huang}, {Jha}, {Johnson}, {Kartaltepe},
  {Khostovan}, {Kirshner}, {Lani}, {Lee}, {Li}, {Madau}, {McCarthy},
  {McIntosh}, {McLure}, {McPartland}, {Mobasher}, {Moreira}, {Mortlock},
  {Moustakas}, {Mozena}, {Nandra}, {Newman}, {Nielsen}, {Niemi}, {Noeske},
  {Papovich}, {Pentericci}, {Pope}, {Primack}, {Ravindranath}, {Reddy},
  {Renzini}, {Rix}, {Robaina}, {Rosario}, {Rosati}, {Salimbeni}, {Scarlata},
  {Siana}, {Simard}, {Smidt}, {Snyder}, {Somerville}, {Spinrad}, {Straughn},
  {Telford}, {Teplitz}, {Trump}, {Vargas}, {Villforth}, {Wagner}, {Wandro},
  {Wechsler}, {Weiner}, {Wiklind}, {Wild}, {Wilson}, {Wuyts}, \&
  {Yun}}]{Koekemoer11}
{Koekemoer} A.~M., {Faber} S.~M., {Ferguson} H.~C., {Grogin} N.~A., {Kocevski}
  D.~D., {Koo} D.~C., {Lai} K., {Lotz} J.~M., {Lucas} R.~A., {McGrath} E.~J.,
  {Ogaz} S., {Rajan} A., {Riess} A.~G., {Rodney} S.~A., {Strolger} L.,
  {Casertano} S., {Castellano} M., {Dahlen} T., {Dickinson} M., {Dolch} T.,
  {Fontana} A., {Giavalisco} M., {Grazian} A., {Guo} Y., {Hathi} N.~P., {Huang}
  K.-H., {van der Wel} A., {Yan} H.-J., {Acquaviva} V., {Alexander} D.~M.,
  {Almaini} O., {Ashby} M.~L.~N., {Barden} M., {Bell} E.~F., {Bournaud} F.,
  {Brown} T.~M., {Caputi} K.~I., {Cassata} P., {Challis} P.~J., {Chary} R.-R.,
  {Cheung} E., {Cirasuolo} M., {Conselice} C.~J., {Roshan Cooray} A., {Croton}
  D.~J., {Daddi} E., {Dav{\'e}} R., {de Mello} D.~F., {de Ravel} L., {Dekel}
  A., {Donley} J.~L., {Dunlop} J.~S., {Dutton} A.~A., {Elbaz} D., {Fazio}
  G.~G., {Filippenko} A.~V., {Finkelstein} S.~L., {Frazer} C., {Gardner} J.~P.,
  {Garnavich} P.~M., {Gawiser} E., {Gruetzbauch} R., {Hartley} W.~G.,
  {H{\"a}ussler} B., {Herrington} J., {Hopkins} P.~F., {Huang} J.-S., {Jha}
  S.~W., {Johnson} A., {Kartaltepe} J.~S., {Khostovan} A.~A., {Kirshner} R.~P.,
  {Lani} C., {Lee} K.-S., {Li} W., {Madau} P., {McCarthy} P.~J., {McIntosh}
  D.~H., {McLure} R.~J., {McPartland} C., {Mobasher} B., {Moreira} H.,
  {Mortlock} A., {Moustakas} L.~A., {Mozena} M., {Nandra} K., {Newman} J.~A.,
  {Nielsen} J.~L., {Niemi} S., {Noeske} K.~G., {Papovich} C.~J., {Pentericci}
  L., {Pope} A., {Primack} J.~R., {Ravindranath} S., {Reddy} N.~A., {Renzini}
  A., {Rix} H.-W., {Robaina} A.~R., {Rosario} D.~J., {Rosati} P., {Salimbeni}
  S., {Scarlata} C., {Siana} B., {Simard} L., {Smidt} J., {Snyder} D.,
  {Somerville} R.~S., {Spinrad} H., {Straughn} A.~N., {Telford} O., {Teplitz}
  H.~I., {Trump} J.~R., {Vargas} C., {Villforth} C., {Wagner} C.~R., {Wandro}
  P., {Wechsler} R.~H., {Weiner} B.~J., {Wiklind} T., {Wild} V., {Wilson} G.,
  {Wuyts} S., {Yun} M.~S., 2011, \apjs, 197, 36

\bibitem[{{Kron}(1980)}]{Kron80}
{Kron} R.~G., 1980, \apjs, 43, 305

\bibitem[{{Lani} {et~al.}(2013){Lani}, {Almaini}, {Hartley}, {Mortlock},
  {H{\"a}u{\ss}ler}, {Chuter}, {Simpson}, {van der Wel}, {Gr{\"u}tzbauch},
  {Conselice}, {Bradshaw}, {Cooper}, {Faber}, {Grogin}, {Kocevski},
  {Koekemoer}, \& {Lai}}]{Lani13}
{Lani} C., {Almaini} O., {Hartley} W.~G., {Mortlock} A., {H{\"a}u{\ss}ler} B.,
  {Chuter} R.~W., {Simpson} C., {van der Wel} A., {Gr{\"u}tzbauch} R.,
  {Conselice} C.~J., {Bradshaw} E.~J., {Cooper} M.~C., {Faber} S.~M., {Grogin}
  N.~A., {Kocevski} D.~D., {Koekemoer} A.~M., {Lai} K., 2013, \mnras, 435, 207

\bibitem[{{Laporte} {et~al.}(2013){Laporte}, {White}, {Naab}, \&
  {Gao}}]{Laporte13}
{Laporte} C.~F.~P., {White} S.~D.~M., {Naab} T., {Gao} L., 2013, \mnras, 435,
  901

\bibitem[{{Laporte} {et~al.}(2012){Laporte}, {White}, {Naab}, {Ruszkowski}, \&
  {Springel}}]{Laporte12}
{Laporte} C.~F.~P., {White} S.~D.~M., {Naab} T., {Ruszkowski} M., {Springel}
  V., 2012, \mnras, 424, 747

\bibitem[{{Lawrence} {et~al.}(2007){Lawrence}, {Warren}, {Almaini}, {Edge},
  {Hambly}, {Jameson}, {Lucas}, {Casali}, {Adamson}, {Dye}, {Emerson},
  {Foucaud}, {Hewett}, {Hirst}, {Hodgkin}, {Irwin}, {Lodieu}, {McMahon},
  {Simpson}, {Smail}, {Mortlock}, \& {Folger}}]{Lawrence07}
{Lawrence} A., {Warren} S.~J., {Almaini} O., {Edge} A.~C., {Hambly} N.~C.,
  {Jameson} R.~F., {Lucas} P., {Casali} M., {Adamson} A., {Dye} S., {Emerson}
  J.~P., {Foucaud} S., {Hewett} P., {Hirst} P., {Hodgkin} S.~T., {Irwin} M.~J.,
  {Lodieu} N., {McMahon} R.~G., {Simpson} C., {Smail} I., {Mortlock} D.,
  {Folger} M., 2007, \mnras, 379, 1599

\bibitem[{{Leja} {et~al.}(2013){Leja}, {van Dokkum}, \& {Franx}}]{Leja13}
{Leja} J., {van Dokkum} P., {Franx} M., 2013, \apj, 766, 33

\bibitem[{{Lidman} {et~al.}(2013){Lidman}, {Iacobuta}, {Bauer}, {Barrientos},
  {Cerulo}, {Couch}, {Delaye}, {Demarco}, {Ellingson}, {Faloon}, {Gilbank},
  {Huertas-Company}, {Mei}, {Meyers}, {Muzzin}, {Noble}, {Nantais}, {Rettura},
  {Rosati}, {S{\'a}nchez-Janssen}, {Strazzullo}, {Webb}, {Wilson}, {Yan}, \&
  {Yee}}]{Lidman13}
{Lidman} C., {Iacobuta} G., {Bauer} A.~E., {Barrientos} L.~F., {Cerulo} P.,
  {Couch} W.~J., {Delaye} L., {Demarco} R., {Ellingson} E., {Faloon} A.~J.,
  {Gilbank} D., {Huertas-Company} M., {Mei} S., {Meyers} J., {Muzzin} A.,
  {Noble} A., {Nantais} J., {Rettura} A., {Rosati} P., {S{\'a}nchez-Janssen}
  R., {Strazzullo} V., {Webb} T.~M.~A., {Wilson} G., {Yan} R., {Yee} H.~K.~C.,
  2013, \mnras, 433, 825

\bibitem[{{Lidman} {et~al.}(2012){Lidman}, {Suherli}, {Muzzin}, {Wilson},
  {Demarco}, {Brough}, {Rettura}, {Cox}, {DeGroot}, {Yee}, {Gilbank},
  {Hoekstra}, {Balogh}, {Ellingson}, {Hicks}, {Nantais}, {Noble}, {Lacy},
  {Surace}, \& {Webb}}]{Lidman12}
{Lidman} C., {Suherli} J., {Muzzin} A., {Wilson} G., {Demarco} R., {Brough} S.,
  {Rettura} A., {Cox} J., {DeGroot} A., {Yee} H.~K.~C., {Gilbank} D.,
  {Hoekstra} H., {Balogh} M., {Ellingson} E., {Hicks} A., {Nantais} J., {Noble}
  A., {Lacy} M., {Surace} J., {Webb} T., 2012, \mnras, 427, 550

\bibitem[{{Lin} {et~al.}(2013){Lin}, {Brodwin}, {Gonzalez}, {Bode},
  {Eisenhardt}, {Stanford}, \& {Vikhlinin}}]{Lin13}
{Lin} Y.-T., {Brodwin} M., {Gonzalez} A.~H., {Bode} P., {Eisenhardt} P.~R.~M.,
  {Stanford} S.~A., {Vikhlinin} A., 2013, \apj, 771, 61

\bibitem[{{Liu} {et~al.}(2009){Liu}, {Mao}, {Deng}, {Xia}, \& {Wen}}]{Liu09}
{Liu} F.~S., {Mao} S., {Deng} Z.~G., {Xia} X.~Y., {Wen} Z.~L., 2009, \mnras,
  396, 2003

\bibitem[{{Marchesini} {et~al.}(2014){Marchesini}, {Muzzin}, {Stefanon},
  {Franx}, {Brammer}, {Marsan}, {Vulcani}, {Fynbo}, {Milvang-Jensen}, {Dunlop},
  \& {Buitrago}}]{Marchesini14}
{Marchesini} D., {Muzzin} A., {Stefanon} M., {Franx} M., {Brammer} G.~G.,
  {Marsan} C.~Z., {Vulcani} B., {Fynbo} J.~P.~U., {Milvang-Jensen} B., {Dunlop}
  J.~S., {Buitrago} F., 2014, \apj, 794, 65

\bibitem[{{Miller} {et~al.}(2005){Miller}, {Nichol}, {Reichart}, {Wechsler},
  {Evrard}, {Annis}, {McKay}, {Bahcall}, {Bernardi}, {Boehringer}, {Connolly},
  {Goto}, {Kniazev}, {Lamb}, {Postman}, {Schneider}, {Sheth}, \&
  {Voges}}]{Miller05}
{Miller} C.~J., {Nichol} R.~C., {Reichart} D., {Wechsler} R.~H., {Evrard}
  A.~E., {Annis} J., {McKay} T.~A., {Bahcall} N.~A., {Bernardi} M.,
  {Boehringer} H., {Connolly} A.~J., {Goto} T., {Kniazev} A., {Lamb} D.,
  {Postman} M., {Schneider} D.~P., {Sheth} R.~K., {Voges} W., 2005, \aj, 130,
  968

\bibitem[{{Mo} {et~al.}(1998){Mo}, {Mao}, \& {White}}]{Mo98}
{Mo} H.~J., {Mao} S., {White} S.~D.~M., 1998, \mnras, 295, 319

\bibitem[{{Mortlock} {et~al.}(2015){Mortlock}, {Conselice}, {Hartley},
  {Duncan}, {Lani}, {Ownsworth}, {Almaini}, {Wel}, {Huang}, {Ashby}, {Willner},
  {Fontana}, {Dekel}, {Koekemoer}, {Ferguson}, {Faber}, {Grogin}, \&
  {Kocevski}}]{Mortlock15}
{Mortlock} A., {Conselice} C.~J., {Hartley} W.~G., {Duncan} K., {Lani} C.,
  {Ownsworth} J.~R., {Almaini} O., {Wel} A.~v.~d., {Huang} K.-H., {Ashby}
  M.~L.~N., {Willner} S.~P., {Fontana} A., {Dekel} A., {Koekemoer} A.~M.,
  {Ferguson} H.~C., {Faber} S.~M., {Grogin} N.~A., {Kocevski} D.~D., 2015,
  \mnras, 447, 2

\bibitem[{{Mortlock} {et~al.}(2013){Mortlock}, {Conselice}, {Hartley},
  {Ownsworth}, {Lani}, {Bluck}, {Almaini}, {Duncan}, {van der Wel},
  {Koekemoer}, {Dekel}, {Dav{\'e}}, {Ferguson}, {de Mello}, {Newman}, {Faber},
  {Grogin}, {Kocevski}, \& {Lai}}]{Mortlock13}
{Mortlock} A., {Conselice} C.~J., {Hartley} W.~G., {Ownsworth} J.~R., {Lani}
  C., {Bluck} A.~F.~L., {Almaini} O., {Duncan} K., {van der Wel} A.,
  {Koekemoer} A.~M., {Dekel} A., {Dav{\'e}} R., {Ferguson} H.~C., {de Mello}
  D.~F., {Newman} J.~A., {Faber} S.~M., {Grogin} N.~A., {Kocevski} D.~D., {Lai}
  K., 2013, \mnras, 433, 1185

\bibitem[{{Muldrew} {et~al.}(2011){Muldrew}, {Pearce}, \& {Power}}]{Muldrew11}
{Muldrew} S.~I., {Pearce} F.~R., {Power} C., 2011, \mnras, 410, 2617

\bibitem[{{Mundy} {et~al.}(2015){Mundy}, {Conselice}, \& {Ownsworth}}]{Mundy15}
{Mundy} C.~J., {Conselice} C.~J., {Ownsworth} J.~R., 2015, \mnras, 450, 3696

\bibitem[{{Naab} {et~al.}(2009){Naab}, {Johansson}, \& {Ostriker}}]{Naab09}
{Naab} T., {Johansson} P.~H., {Ostriker} J.~P., 2009, \apjl, 699, L178

\bibitem[{{Ownsworth} {et~al.}(2014){Ownsworth}, {Conselice}, {Mortlock},
  {Hartley}, {Almaini}, {Duncan}, \& {Mundy}}]{Ownsworth14}
{Ownsworth} J.~R., {Conselice} C.~J., {Mortlock} A., {Hartley} W.~G., {Almaini}
  O., {Duncan} K., {Mundy} C.~J., 2014, \mnras, 445, 2198

\bibitem[{{Ownsworth} {et~al.}(2012){Ownsworth}, {Conselice}, {Mortlock},
  {Hartley}, \& {Buitrago}}]{Ownsworth12}
{Ownsworth} J.~R., {Conselice} C.~J., {Mortlock} A., {Hartley} W.~G.,
  {Buitrago} F., 2012, \mnras, 426, 764

\bibitem[{{Papovich} {et~al.}(2006){Papovich}, {Moustakas}, {Dickinson}, {Le
  Floc'h}, {Rieke}, {Daddi}, {Alexander}, {Bauer}, {Brandt}, {Dahlen}, {Egami},
  {Eisenhardt}, {Elbaz}, {Ferguson}, {Giavalisco}, {Lucas}, {Mobasher},
  {P{\'e}rez-Gonz{\'a}lez}, {Stutz}, {Rieke}, \& {Yan}}]{Papovich06}
{Papovich} C., {Moustakas} L.~A., {Dickinson} M., {Le Floc'h} E., {Rieke}
  G.~H., {Daddi} E., {Alexander} D.~M., {Bauer} F., {Brandt} W.~N., {Dahlen}
  T., {Egami} E., {Eisenhardt} P., {Elbaz} D., {Ferguson} H.~C., {Giavalisco}
  M., {Lucas} R.~A., {Mobasher} B., {P{\'e}rez-Gonz{\'a}lez} P.~G., {Stutz} A.,
  {Rieke} M.~J., {Yan} H., 2006, \apj, 640, 92

\bibitem[{{Peng} {et~al.}(2002){Peng}, {Ho}, {Impey}, \& {Rix}}]{Peng02}
{Peng} C.~Y., {Ho} L.~C., {Impey} C.~D., {Rix} H.-W., 2002, \aj, 124, 266

\bibitem[{{Salim} {et~al.}(2007){Salim}, {Rich}, {Charlot}, {Brinchmann},
  {Johnson}, {Schiminovich}, {Seibert}, {Mallery}, {Heckman}, {Forster},
  {Friedman}, {Martin}, {Morrissey}, {Neff}, {Small}, {Wyder}, {Bianchi},
  {Donas}, {Lee}, {Madore}, {Milliard}, {Szalay}, {Welsh}, \& {Yi}}]{Salim07}
{Salim} S., {Rich} R.~M., {Charlot} S., {Brinchmann} J., {Johnson} B.~D.,
  {Schiminovich} D., {Seibert} M., {Mallery} R., {Heckman} T.~M., {Forster} K.,
  {Friedman} P.~G., {Martin} D.~C., {Morrissey} P., {Neff} S.~G., {Small} T.,
  {Wyder} T.~K., {Bianchi} L., {Donas} J., {Lee} Y.-W., {Madore} B.~F.,
  {Milliard} B., {Szalay} A.~S., {Welsh} B.~Y., {Yi} S.~K., 2007, \apjs, 173,
  267

\bibitem[{{S{\'e}rsic}(1963)}]{Sersic63}
{S{\'e}rsic} J.~L., 1963, Boletin de la Asociacion Argentina de Astronomia La
  Plata Argentina, 6, 41

\bibitem[{{Shankar} {et~al.}(2015){Shankar}, {Buchan}, {Rettura}, {Bouillot},
  {Moreno}, {Licitra}, {Bernardi}, {Huertas-Company}, {Mei}, {Ascaso}, {Sheth},
  {Delaye}, \& {Raichoor}}]{Shankar15}
{Shankar} F., {Buchan} S., {Rettura} A., {Bouillot} V.~R., {Moreno} J.,
  {Licitra} R., {Bernardi} M., {Huertas-Company} M., {Mei} S., {Ascaso} B.,
  {Sheth} R., {Delaye} L., {Raichoor} A., 2015, \apj, 802, 73

\bibitem[{{Trujillo} {et~al.}(2007){Trujillo}, {Conselice}, {Bundy}, {Cooper},
  {Eisenhardt}, \& {Ellis}}]{Trujillo07}
{Trujillo} I., {Conselice} C.~J., {Bundy} K., {Cooper} M.~C., {Eisenhardt} P.,
  {Ellis} R.~S., 2007, \mnras, 382, 109

\bibitem[{{Twite} {et~al.}(2012){Twite}, {Conselice}, {Buitrago}, {Noeske},
  {Weiner}, {Acosta-Pulido}, \& {Bauer}}]{Twite12}
{Twite} J.~W., {Conselice} C.~J., {Buitrago} F., {Noeske} K., {Weiner} B.~J.,
  {Acosta-Pulido} J.~A., {Bauer} A.~E., 2012, \mnras, 420, 1061

\bibitem[{{van der Wel} {et~al.}(2014){van der Wel}, {Franx}, {van Dokkum},
  {Skelton}, {Momcheva}, {Whitaker}, {Brammer}, {Bell}, {Rix}, {Wuyts},
  {Ferguson}, {Holden}, {Barro}, {Koekemoer}, {Chang}, {McGrath},
  {H{\"a}ussler}, {Dekel}, {Behroozi}, {Fumagalli}, {Leja}, {Lundgren},
  {Maseda}, {Nelson}, {Wake}, {Patel}, {Labb{\'e}}, {Faber}, {Grogin}, \&
  {Kocevski}}]{vanderWel14}
{van der Wel} A., {Franx} M., {van Dokkum} P.~G., {Skelton} R.~E., {Momcheva}
  I.~G., {Whitaker} K.~E., {Brammer} G.~B., {Bell} E.~F., {Rix} H.-W., {Wuyts}
  S., {Ferguson} H.~C., {Holden} B.~P., {Barro} G., {Koekemoer} A.~M., {Chang}
  Y.-Y., {McGrath} E.~J., {H{\"a}ussler} B., {Dekel} A., {Behroozi} P.,
  {Fumagalli} M., {Leja} J., {Lundgren} B.~F., {Maseda} M.~V., {Nelson} E.~J.,
  {Wake} D.~A., {Patel} S.~G., {Labb{\'e}} I., {Faber} S.~M., {Grogin} N.~A.,
  {Kocevski} D.~D., 2014, \apj, 788, 28

\bibitem[{{van Dokkum} {et~al.}(2004){van Dokkum}, {Franx}, {F{\"o}rster
  Schreiber}, {Illingworth}, {Daddi}, {Knudsen}, {Labb{\'e}}, {Moorwood},
  {Rix}, {R{\"o}ttgering}, {Rudnick}, {Trujillo}, {van der Werf}, {van der
  Wel}, {van Starkenburg}, \& {Wuyts}}]{vanDokkum04}
{van Dokkum} P.~G., {Franx} M., {F{\"o}rster Schreiber} N.~M., {Illingworth}
  G.~D., {Daddi} E., {Knudsen} K.~K., {Labb{\'e}} I., {Moorwood} A., {Rix}
  H.-W., {R{\"o}ttgering} H., {Rudnick} G., {Trujillo} I., {van der Werf} P.,
  {van der Wel} A., {van Starkenburg} L., {Wuyts} S., 2004, \apj, 611, 703

\bibitem[{{van Dokkum} {et~al.}(2010){van Dokkum}, {Whitaker}, {Brammer},
  {Franx}, {Kriek}, {Labb{\'e}}, {Marchesini}, {Quadri}, {Bezanson},
  {Illingworth}, {Muzzin}, {Rudnick}, {Tal}, \& {Wake}}]{vanDokkum10}
{van Dokkum} P.~G., {Whitaker} K.~E., {Brammer} G., {Franx} M., {Kriek} M.,
  {Labb{\'e}} I., {Marchesini} D., {Quadri} R., {Bezanson} R., {Illingworth}
  G.~D., {Muzzin} A., {Rudnick} G., {Tal} T., {Wake} D., 2010, \apj, 709, 1018

\bibitem[{{von der Linden} {et~al.}(2007){von der Linden}, {Best}, {Kauffmann},
  \& {White}}]{Linden07}
{von der Linden} A., {Best} P.~N., {Kauffmann} G., {White} S.~D.~M., 2007,
  \mnras, 379, 867

\bibitem[{{Whiley} {et~al.}(2008){Whiley}, {Arag{\'o}n-Salamanca}, {De Lucia},
  {von der Linden}, {Bamford}, {Best}, {Bremer}, {Jablonka}, {Johnson},
  {Milvang-Jensen}, {Noll}, {Poggianti}, {Rudnick}, {Saglia}, {White}, \&
  {Zaritsky}}]{Whiley08}
{Whiley} I.~M., {Arag{\'o}n-Salamanca} A., {De Lucia} G., {von der Linden} A.,
  {Bamford} S.~P., {Best} P., {Bremer} M.~N., {Jablonka} P., {Johnson} O.,
  {Milvang-Jensen} B., {Noll} S., {Poggianti} B.~M., {Rudnick} G., {Saglia} R.,
  {White} S., {Zaritsky} D., 2008, \mnras, 387, 1253

\bibitem[{{Zhang} {et~al.}(2016){Zhang}, {Miller}, {McKay}, {Rooney}, {Evrard},
  {Romer}, {Perfecto}, {Song}, {Desai}, {Mohr}, {Wilcox}, {Bermeo-Hernandez},
  {Jeltema}, {Hollowood}, {Bacon}, {Capozzi}, {Collins}, {Das}, {Gerdes},
  {Hennig}, {Hilton}, {Hoyle}, {Kay}, {Liddle}, {Mann}, {Mehrtens}, {Nichol},
  {Papovich}, {Sahl{\'e}n}, {Soares-Santos}, {Stott}, {Viana}, {Abbott},
  {Abdalla}, {Banerji}, {Bauer}, {Benoit-L{\'e}vy}, {Bertin}, {Brooks},
  {Buckley-Geer}, {Burke}, {Carnero Rosell}, {Castander}, {Diehl}, {Doel},
  {Cunha}, {Eifler}, {Fausti Neto}, {Fernandez}, {Flaugher}, {Fosalba},
  {Frieman}, {Gaztanaga}, {Gruen}, {Gruendl}, {Honscheid}, {James}, {Kuehn},
  {Kuropatkin}, {Lahav}, {Maia}, {Makler}, {Marshall}, {Martini}, {Miquel},
  {Ogando}, {Plazas}, {Roodman}, {Rykoff}, {Sako}, {Sanchez}, {Scarpine},
  {Schubnell}, {Sevilla}, {Smith}, {Sobreira}, {Suchyta}, {Swanson}, {Tarle},
  {Thaler}, {Tucker}, {Vikram}, \& {da Costa}}]{Zhang15}
{Zhang} Y., {Miller} C., {McKay} T., {Rooney} P., {Evrard} A.~E., {Romer}
  A.~K., {Perfecto} R., {Song} J., {Desai} S., {Mohr} J., {Wilcox} H.,
  {Bermeo-Hernandez} A., {Jeltema} T., {Hollowood} D., {Bacon} D., {Capozzi}
  D., {Collins} C., {Das} R., {Gerdes} D., {Hennig} C., {Hilton} M., {Hoyle}
  B., {Kay} S., {Liddle} A., {Mann} R.~G., {Mehrtens} N., {Nichol} R.~C.,
  {Papovich} C., {Sahl{\'e}n} M., {Soares-Santos} M., {Stott} J., {Viana}
  P.~T., {Abbott} T., {Abdalla} F.~B., {Banerji} M., {Bauer} A.~H.,
  {Benoit-L{\'e}vy} A., {Bertin} E., {Brooks} D., {Buckley-Geer} E., {Burke}
  D.~L., {Carnero Rosell} A., {Castander} F.~J., {Diehl} H.~T., {Doel} P.,
  {Cunha} C.~E., {Eifler} T.~F., {Fausti Neto} A., {Fernandez} E., {Flaugher}
  B., {Fosalba} P., {Frieman} J., {Gaztanaga} E., {Gruen} D., {Gruendl} R.~A.,
  {Honscheid} K., {James} D., {Kuehn} K., {Kuropatkin} N., {Lahav} O., {Maia}
  M.~A.~G., {Makler} M., {Marshall} J.~L., {Martini} P., {Miquel} R., {Ogando}
  R., {Plazas} A.~A., {Roodman} A., {Rykoff} E.~S., {Sako} M., {Sanchez} E.,
  {Scarpine} V., {Schubnell} M., {Sevilla} I., {Smith} R.~C., {Sobreira} F.,
  {Suchyta} E., {Swanson} M.~E.~C., {Tarle} G., {Thaler} J., {Tucker} D.,
  {Vikram} V., {da Costa} L.~N., 2016, \apj, 816, 98

\bibitem[{{Zhao} {et~al.}(2015{\natexlab{a}}){Zhao}, {Arag{\'o}n-Salamanca}, \&
  {Conselice}}]{Zhao15a}
{Zhao} D., {Arag{\'o}n-Salamanca} A., {Conselice} C.~J., 2015{\natexlab{a}},
  \mnras, 453, 4444

\bibitem[{{Zhao} {et~al.}(2015{\natexlab{b}}){Zhao}, {Arag{\'o}n-Salamanca}, \&
  {Conselice}}]{Zhao15b}
---, 2015{\natexlab{b}}, \mnras, 448, 2530

\end{thebibliography}

\label{lastpage}
\end{document}